\documentclass[preprint,3p,10pt]{elsarticle} 

\usepackage{siunitx}
\usepackage[export]{adjustbox}

\usepackage{graphicx}
\usepackage{units}

\usepackage{booktabs}

\usepackage{multirow}

\usepackage{pgfplots}
\usepackage{pgfplotstable}
\usepackage{tikz}
\usepackage{environ}
\pgfplotsset{compat=newest}
\pgfplotsset{plot coordinates/math parser=false}
\usepgfplotslibrary{groupplots,ternary}
\usetikzlibrary{calc,shapes,arrows.meta,backgrounds,arrows,positioning,pgfplots.groupplots,matrix}

\usepgfplotslibrary{external}
\usepgfplotslibrary{groupplots}
\makeatletter
\newsavebox{\measure@tikzpicture}
\NewEnviron{scaletikzpicturetowidth}[1]{%
  \def\tikz@width{#1}%
  \begin{lrbox}{\measure@tikzpicture}%
  \BODY
  \end{lrbox}%
  \pgfmathparse{#1/\wd\measure@tikzpicture}%
  \BODY
}
\makeatother

\usepackage{amsmath}
\usepackage{xcolor}

\definecolor{color1}{RGB}{228,26,28}
\definecolor{color2}{RGB}{55,126,184}
\definecolor{color3}{RGB}{77,175,74}
\definecolor{color4}{RGB}{152,78,163}

\definecolor{colora}{RGB}{228,26,28}
\definecolor{colorb}{RGB}{55,126,184}
\definecolor{colorc}{RGB}{77,175,74}
\definecolor{colord}{RGB}{152,78,163}
\definecolor{colore}{RGB}{255,127,0}

\definecolor{redVLE}{RGB}{213,62,79}
\definecolor{blueVLE}{RGB}{50,136,189}

\newcommand{\at}[2][]{#1|_{#2}}

\pgfplotsset{select coords between index/.style 2 args={
    x filter/.code={
        \ifnum\coordindex<#1\def\pgfmathresult{}\fi
        \ifnum\coordindex>#2\def\pgfmathresult{}\fi
    }
}}

\usetikzlibrary{decorations.markings}

\makeatletter
\def\ps@pprintTitle{%
	\let\@oddhead\@empty
	\let\@evenhead\@empty
	\def\@oddfoot{}%
	\let\@evenfoot\@oddfoot}
\makeatother

\usepackage{setspace}

\usepackage{natbib}

\begin{document}

\begin{frontmatter}
	
	\vspace*{60pt}
	  \title{\textbf{Real-Gas Effects and Phase Separation in\\ Underexpanded Jets at Engine-Relevant Conditions}\tnoteref{t1}}
	
	\tnotetext[t1]{Preprint submitted to AIAA Scitech 2018, Kissimmee, Florida}
	
	\author[UniBw]{Christoph Traxinger\corref{cor1}}
	\cortext[cor1]{Corresponding author: christoph.traxinger@unibw.de}
	\author[UniBw]{Matthias Banholzer}
	\author[UniBw]{Michael Pfitzner}
	\address[UniBw]{Institute for Thermodynamics, Bundeswehr University Munich,Werner-Heisenberg-Weg 39, 85577 Neubiberg, Germany}

\begin{abstract}
A numerical framework implemented in the open-source tool OpenFOAM is presented in this work combining a hybrid, pressure-based solver with a vapor-liquid equilibrium model based on the cubic equation of state. This framework is used in the present work to investigate underexpanded jets at engine-relevant conditions where real-gas effects and mixture induced phase separation are probable to occur. A thorough validation and discussion of the applied vapor-liquid equilibrium model is conducted by means of general thermodynamic relations and measurement data available in the literature. Engine-relevant simulation cases for two different fuels were defined. Analyses of the flow field show that the used fuel has a first order effect on the occurrence of phase separation. In the case of phase separation two different effects could be revealed causing the single-phase instability, namely the strong expansion and the mixing of the fuel with the chamber gas. A comparison of single-phase and two-phase jets disclosed that the phase separation leads to a completely different penetration depth in contrast to single-phase injection and therefore commonly used analytical approaches fail to predict the penetration depth.
\end{abstract}

\end{frontmatter}


\section{Introduction}

Many of today's and future transportation and power generation systems are based on the combustion of fossil fuels. In recent years, the concerns of environmental protection and global warming increased significantly. Thus, the need of the reduction of emissions and fuel consumption of engines are the major drivers for new innovations and technical improvements. One of the major trends in all types of engines is the steady increase in operating pressure to fulfill the aforementioned goals. This leads to situations where the mixture and the combustion of fuel and oxidizer takes place at supercritical pressures ($p > p_c$) with respect to the pure components value. At supercritical state, the thermodynamic properties are non-linear functions of temperature and pressure and the widely used ideal gas law is not valid anymore, especially at low/cryogenic and moderate temperatures. This aspect is crucial during the injection process and many researchers have therefore focused on understanding and optimizing the high-pressure injection prior to combustion, mostly using computational fluid \mbox{dynamics (CFD)}.

\begin{figure}[h]
	\centering
	\resizebox{0.5\textwidth}{!}{%
	\tikzsetnextfilename{underexpandedJet_schematic}%
	\begin{tikzpicture}[scale=0.60]
		\tiny
		\draw (0,0) -- (0,3);
		\draw (-1,0.5) -- (0,0.5);
		\draw [dashdotted] (0,0) -- (8,0);
		\draw  (0,0.5) to[out=35,in=175] (6,2.0);
		\draw  (0,0.5) to[out=45,in=175] (7.5,3.0);
		\draw  (6,2.0) to[out=60,in=215] (7.5,3.0);
		\draw  (6,2.0) to[out=0,in=175] (7.5,1.9);
		\draw  (6,2.0) to[out=-85,in=90] (6.15,0.0);
		\draw  (2,2.055) to[out=0,in=175] (6,2.0);
		\draw  (2,2.055) -- (0,0.5);
		\draw  (1,1.385) to[out=17.5,in=175] (6,2.0);
		\draw  (1,1.385) -- (0,0.5);
		\draw  (0,0.5) -- (1,0.85);
		\draw  (0,0.5) -- (0.8,0.7);
		\draw  (0,0.5) -- (1,0.4);
		\draw  (0,0.5) -- (0.9,0.25);
		
		\draw  (7.5,3.0) -- (7.7,3.1);
		\draw  (7.5,3.0) -- (7.7,2.9);
		\draw  (7.5,3.0) -- (7.5,2.8);
		\draw  (7.5,3.0) -- (7.575,2.8);
		
		\begin{scope}[yscale=-1,xscale=1]
			\draw (0,0) -- (0,3);
			\draw (-1,0.5) -- (0,0.5);
			\draw  (0,0.5) to[out=35,in=175] (6,2.0);
			\draw  (0,0.5) to[out=45,in=175] (7.5,3.0);
			\draw  (6,2.0) to[out=60,in=215] (7.5,3.0);
			\draw  (6,2.0) to[out=0,in=175] (7.5,1.9);
			\draw  (6,2.0) to[out=-85,in=90] (6.15,0.0);
			\draw  (2,2.055) to[out=0,in=175] (6,2.0);
			\draw  (2,2.055) -- (0,0.5);
			\draw  (1,1.385) to[out=17.5,in=175] (6,2.0);
			\draw  (1,1.385) -- (0,0.5);
			\draw  (0,0.5) -- (1,0.85);
			\draw  (0,0.5) -- (0.8,0.7);
			\draw  (0,0.5) -- (1,0.4);
			\draw  (0,0.5) -- (0.9,0.25);
			
			\draw  (7.5,3.0) -- (7.7,3.1);
			\draw  (7.5,3.0) -- (7.7,2.9);
			\draw  (7.5,3.0) -- (7.5,2.8);
			\draw  (7.5,3.0) -- (7.575,2.8);
		\end{scope}
		
		\draw [<-,thick] (6.2,1.4) -- (6.7,1.5) node[anchor=west] {Mach disk};
		\draw [<-,thick] (6.5,-1.9) -- (6.7,-1.5) node[anchor=west] {Slip line};
		\draw [<-,thick] (6.5,2.4) -- (6.9,2.3) node[anchor=west] {Reflected shock};
		\draw [<-,thick] (1.65,1.65) -- (1.55,2.5) node[anchor=south,align=center] {Compression \\ waves};
		\draw [<-,thick] (1.5,1.2) -- (2.2,1.0) node[anchor=west,align=left] {Intercepting \\ shock};
		\draw [<-,thick] (2.3,-2.3) -- (1.75,-2.75) node[anchor=north] {Jet boundary};
		\node at (1.0,-0.6) [anchor=west,align=left] {Expansion \\ fan};
		
		\node at (0,0) [anchor=east] {Ma $= 1$};
		\node at (4,2.5) [anchor=west] {Ma $> 1$};
		\node at (6.0,-0.7) [anchor=east] {Ma $\gg 1$};
		\node at (6.2,-0.7) [anchor=west] {Ma $< 1$};
		
		\end{tikzpicture}%

	}
	\caption{Schematic of an underexpanded jet showing a series of expansions and shock structures~\cite{crist1966study}.}
	\label{fig:cristShock}
\end{figure}
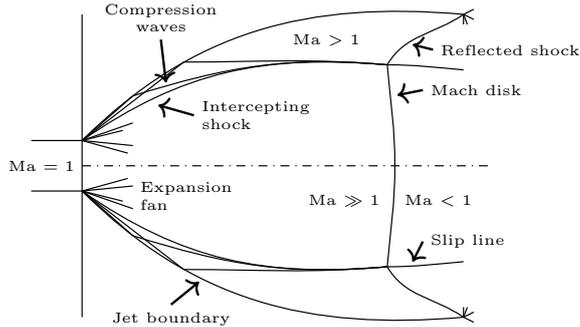

In the field of liquid or liquid-like fuel injection into a gaseous environment at supercritical pressure typically encountered in gasoline, diesel and rocket engines, it is state-of-the-art to conduct large-eddy simulations (LES) based on real-gas thermodynamics. Pioneer work was done by Oefelein and Yang~\cite{oefelein1998a} as well as Zong \textit{et al.}~\cite{zong2004a}. Many research groups, e.g., Schmitt \textit{et al.}~\cite{schmitt2009a} and M{\"u}ller \textit{et al.}~\cite{muller2016a}, have followed their method and used the dense-gas approach to do LES at rocket engine relevant conditions. More recently, Matheis and Hickel~\cite{matheis2016a} and Traxinger \textit{et al.}~\cite{traxinger2017a} applied a vapor-liquid equilibrium (VLE) model and showed that phase separation due to non-linear mixing phenomena is likely to occur at typical injection conditions of diesel and rocket engines. In contrast to these thoroughly investigated applications, the injection of gaseous fuel, especially at conditions typical for piston engines, is a topic which has not received as much attention as the aforementioned applications. This is changing considerably since the last few years as environmental and global warming aspects are getting more and more important and new fuel concepts have to be considered also for piston engines. Challenges resulting from the direct injection of gaseous fuels at high pressure ratios are the gas dynamics of the near-nozzle flow structure, the gaseous jet penetration and the fuel-oxidizer mixing. These fluid dynamic aspects are accompanied by strong pressure and temperature changes leading to non-negligible real-gas effects, see, e.g., Khaksarfard \textit{et al.}~\cite{khaksarfard2010numerical} and Bonelli \textit{et al.}~\cite{bonelli2013numerical}. In Fig.~\ref{fig:cristShock} the schematic of a highly underexpanded jet~\cite{crist1966study} forming a series of shock and expansion structures is illustrated like it occurs during the injection process when high nozzle pressure ratios are present as it will be the case in this study.\\

Recently, Banholzer~\textit{et al.}~\cite{banholzer2017} presented first results of the injection of gaseous hydrogen into air with the focus on the jet penetration depth as well as the Mach disk position. They showed an excellent agreement between experimental Schlieren images and their simulations. The present study is expanding these results focusing on the comparison of two different fuels, namely a hydrogen and a methane based one, and on the thermodynamics itself because they are a crucial aspect at engine-relevant conditions. We will extend the real-gas framework presented by Banholzer~\textit{et al.}~\cite{banholzer2017} with the VLE-model used in Traxinger~\textit{et al.}~\cite{traxinger2017a} and show that phase separation is likely to occur at engine-relevant conditions. To the authors knowledge this is the first time that simulations of underexpanded jets at such conditions are presented. As experimental investigations in terms of underexpanded jets at real-gas conditions are lacking we will strongly focus on a thorough discussion and validation of the VLE-model by using available experimental data from VLE measurements of the appropriate mixtures. The paper is structured as follows. In the sections \ref{sec:NumericalMethod} and \ref{sec:Thermodynamics} the numerical approach and the VLE model are discussed in detail. Section~\ref{sec:ILASS} presents the validation of the numerical framework (solver and thermodynamic model) using experimental and numerical findings recently presented by Traxinger~\textit{et al.}~\cite{traxinger2017a}. In section~\ref{sec:testCases} the engine-relevant simulation cases for two different fuels are defined and a-priori analyses are carried out concerning phase separation probability and choked nozzle conditions. The main results are discussed in section~\ref{sec:results} while the conclusions and some aspects concerning future work will be outlined in section~\ref{sec:conclusion}.

\section{Numerical Method}
\label{sec:NumericalMethod}

\subsection{Governing Equations}

Numerical simulations were carried out based on the inert and compressible conservation equations for mass, momentum and energy and the transport equation for species $k$
\begin{align}
	\frac{\partial \rho}{\partial t} + \frac{\partial \left(\rho u_i \right)}{\partial \zeta_i} & = 0, \label{eq:NavierStokes1}\\
	\frac{\partial \left(\rho u_i \right)}{\partial t} + \frac{\partial \left(\rho u_i u_j \right)}{\partial \zeta_j} & = -\frac{\partial p}{\partial \zeta_i} + \frac{\partial \sigma_{ij}}{\partial \zeta_j}, \label{eq:NavierStokes2}\\
	\frac{\partial \rho e_\textrm{t}}{\partial t} + \frac{\partial \left(\rho u_i e_\textrm{t} \right)}{\partial \zeta_i} & = - \frac{\partial \left( u_i p \right)}{\partial \zeta_i} + \frac{\partial \left( u_i \sigma_{ij} \right)}{\partial \zeta_j} - \frac{\partial q_i}{\partial \zeta_i}, \label{eq:NavierStokes3}\\
	\frac{\partial \left(\rho Y_k \right)}{\partial t} + \frac{\partial \left(\rho u_i Y_k \right)}{\partial \zeta_i} & = -\frac{\partial D_{k,i}}{\partial \zeta_i}, \label{eq:TransportSpecie}
\end{align}
wherein $t$ is the time, $\mathbf{\zeta}$ is the vector of Cartesian coordinates, $\rho$ is the density, $\mathbf{u}$ is the velocity vector, $p$ is the static pressure, $\sigma_{ij}$ is a component of the viscous stress tensor, $\mathbf{q}$ is the heat flux vector and $e_t$ is the total energy given as $e_t = e + \mathbf{u}^2/2$, where $e$ is the internal energy. Furthermore, $\mathbf{Y}$ is the vector of the mass fractions and $\mathbf{D}$ is the mass flux vector. The heat flux vector and the mass flux vector are modeled using Fourier's and Fick's law, respectively. In the numerical simulations the Reynolds-Averaged Navier-Stokes (RANS) equations are solved. The turbulence closure was achieved with the widely-used SST turbulence model. The final closure problem in terms of primitive and derived thermodynamic properties was solved by applying real-gas thermodynamics, which will be thoroughly discussed in section~\ref{sec:Thermodynamics}. 

\subsection{Hybrid, pressure-based solver}

In order to solve the system of equations a hybrid, pressure-based solver is used which was originally derived by Kraposhin \textit{et~al.}~\cite{kraposhin2015adaptation}. It combines the Pressure Implicit with Splitting of Operator (PISO) algorithm proposed by Issa~\cite{issa1986solution} and the Kurganov-Tadmor~\cite{kurganov2000new} (KT) scheme. While the PISO algorithm is suitable for fluid flows with small Mach numbers, it lacks of accuracy for high speed flows due to the occurrence of numerical oscillations in regions with discontinuities. The KT scheme however is a high-resolution central scheme for trans- and supersonic flows. A blending function based on the local Mach number and the Courant-Friedrichs-Lewy (CFL) criterion switches between the incompressible and compressible flux formulations. Wherever the flow approaches trans- and supersonic flow regimes the blending function manages the switch to the high-resolution central scheme of Kurganov-Tadmor. This scheme uses more precise information of local characteristic propagation speeds at the cell boundaries, which are projected from the cell centers onto the cell faces. This leads to a separate treatment of smooth and non-smooth regions. Non-smooth parts of the computed approximations are averaged over smaller cells than smooth parts. The introduced numerical diffusion is independent of $\Delta t$. As the high-resolution scheme relies on the correct local characteristic propagation speeds at the cell faces, which are dependent on the local speed of sound, the exact evaluation of those is a necessity, for single-phase regions and especially for two-phase regions where the speed of sound drops significantly at the phase boundaries due to inhomogeneities in the fluid and discontinuities become present (see chapter~\ref{sec:Thermodynamics}, section~\ref{subsec:Psi}).\\

Recently, Kraposhin \textit{et~al.}~\cite{kraposhinbanholzer2017} extended the source code of the solver based on the work of Jarczyk and Pfitzner~\cite{jarczyk2012a} and M\"uller \textit{et~al.}~\cite{muller2016b} such that it consistently accounts for real-gas thermodynamics. Additionally, the multi-species transport equations were implemented by Banholzer \textit{et~al.}~\cite{banholzer2017}. In flows with real-gas mixing the thermodynamic properties are much more sensitive to the transported fields, i.e., species, enthalpy and pressure, than in ideal-gas flows. Therefore, it is important to include the species equation, the enthalpy equation as well as the evaluation of the thermodynamic properties into the PISO loop~\cite{jarczyk2012a}. In addition, the modeling of the density $\rho$ in the transient term $\partial \rho/\partial t$ in the pressure equation has to be done differently compared to the ideal-gas approach~\cite{jarczyk2012a}. As the density is not a linear function of the pressure in the real-gas regime a Taylor expansion for the evaluation of the density inside the pressure equation is used
\begin{equation}
\rho = \rho_0 + \frac{\partial \rho}{\partial p}\at[\Bigg]{0} \left( p - p_0\right)
\end{equation}
where the index 0 refers to the base point of the Taylor expansion, i.e., the last time or iteration step. Due to the fact that the transport properties are solved sequentially only the values of the respective transported field change when it is solved. This in turn implies that the enthalpy and the species fields are constant during the evaluation of the pressure equation and therefore $\partial \rho/\partial p$ is evaluated at isenthalpic conditions ($h$ = const.) and at constant compositions ($Y_k$ = const.), i.e., $\partial \rho/\partial p \at[\big]{h,Y_k}$. Therefore, we need to use the isenthalpic compressibility $\psi_h$ as the basis for the evaluation due to the following mathematical relationship:
\begin{equation}
\frac{\partial \rho}{\partial p} \at[\Bigg]{h,Y_k} = - \frac{1}{v^2} \frac{\partial v}{\partial p} \at[\Bigg]{h,Y_k} = \frac{1}{v} \psi_h \; .
\end{equation}
Here, $v$ is the specific volume. The evaluation of the compressibility in single and two-phase systems will be discussed in the next section. For a detailed derivation of the pressure equation and a thorough validation of the solver see Kraposhin \textit{et~al.}~\cite{kraposhinbanholzer2017}.


\section{Thermodynamic Modeling}
\label{sec:Thermodynamics}

\subsection{Real-Gas Modeling}

\begin{figure}[t]
	\centering
	\tikzsetnextfilename{comparisonAlkanes}%

\begin{tikzpicture}
	\begin{groupplot}[group style = {group name = plots,group size = 2 by 1, horizontal sep = 50pt}, width = 0.4\textwidth, height = 0.32\textwidth]
		\nextgroupplot
		[
		xmin=0,
		xmax=11,
		xlabel = Number of carbons,
		xtick={1,2,3,4,5,6,7,8,9,10},
		ymin=350,
		ymax = 650,
		ylabel = $\rho\text{ [kg/m$^3$]}$,
		legend style={legend pos=south east,font=\scriptsize,draw=none}
		]
			\addplot [solid,line width=0.5pt,every mark/.append style={solid, fill=black},mark=*] table [x=Alkane,y=rhoCoolProp]{data/compareEoS__pr_1.5_Tr_0.75.dat};
			\addlegendentry{CoolProp};
			\addplot [dashed,line width=0.5pt,every mark/.append style={solid, fill=white},mark=square*] table [x=Alkane,y=rhoPR]{data/compareEoS__pr_1.5_Tr_0.75.dat};
			\addlegendentry{PR-EoS};
			\addplot [dotted,line width=0.5pt,every mark/.append style={solid, fill=white},mark=diamond*] table [x=Alkane,y=rhoSRK]{data/compareEoS__pr_1.5_Tr_0.75.dat};
			\addlegendentry{SRK-EoS};
			\coordinate (top) at (rel axis cs:0,1);
		\nextgroupplot
		[
		xmin=0,
		xmax=600,
		xlabel = $p \textrm{ [bar]}$,
		x filter/.code={\pgfmathparse{#1*1e-5}\pgfmathresult},
		ymin=0.01,
		ymax=3.5,
		ylabel = {$Z \textrm{ [-]}$},
		legend style={legend pos=north west,font=\scriptsize,draw=none}
		]
			\addplot [only marks,every mark/.append style={solid, fill=black},mark=*,mark repeat=20, mark size=1.0pt] table [x=p,y=ZCoolProp]{data/compareEoS__Fluid_Methane_T_300K.dat};
			\addlegendentry{CoolProp};
			\addplot [dashed,line width=0.5pt] table [x=p,y=ZPR]{data/compareEoS__Fluid_Methane_T_300K.dat};
			\addlegendentry{PR-EoS};
			\addplot [dotted,line width=0.5pt] table [x=p,y=ZSRK]{data/compareEoS__Fluid_Methane_T_300K.dat};
			\addlegendentry{SRK-EoS};
			\node at (axis cs:500,0.5) [anchor=south] {\scriptsize Methane};
			\addplot [only marks,every mark/.append style={solid, fill=black},mark=*,mark repeat=20, mark size=1.0pt] table [x=p,y=ZCoolProp]{data/compareEoS__Fluid_Hexane_T_300K.dat};
			\addplot [dashed,line width=0.5pt] table [x=p,y=ZPR]{data/compareEoS__Fluid_Hexane_T_300K.dat};
			\addplot [dotted,line width=0.5pt] table [x=p,y=ZSRK]{data/compareEoS__Fluid_Hexane_T_300K.dat};
			\node at (axis cs:400,2.5) [anchor=south] {\scriptsize n-Hexane};
			\addplot [only marks,every mark/.append style={solid, fill=black},mark=*,mark repeat=20, mark size=1.0pt] table [x=p,y=ZCoolProp]{data/compareEoS__Fluid_Hydrogen_T_300K.dat};
			\addplot [dashed,line width=0.5pt] table [x=p,y=ZPR]{data/compareEoS__Fluid_Hydrogen_T_300K.dat};
			\addplot [dotted,line width=0.5pt] table [x=p,y=ZSRK]{data/compareEoS__Fluid_Hydrogen_T_300K.dat};
			\node at (axis cs:400,1.25) [anchor=south west] {\scriptsize Hydrogen};
	\end{groupplot}	
	\node[below = 0.4cm of plots c1r1.south,xshift=-3.5cm] {a)};
	\node[below = 0.4cm of plots c2r1.south,xshift=-3.5cm] {b)};
\end{tikzpicture}%

	\caption{Comparison of PR-EoS and SRK-EoS with reference data from CoolProp~\cite{CoolProp}: a) Density $\rho$ for different alkanes from the homologue series at $p_r = \frac{p}{p_\textrm{c}} = 1.5$ and $T_r = \frac{T}{T_\textrm{c}} = 0.75$; b) Compressibility factor $Z = \frac{p v}{\mathcal{R} T}$ of methane, hydrogen and n-hexane at $T$ = 300 K.}
	\label{fig:comparisonAlkanes}
\end{figure}
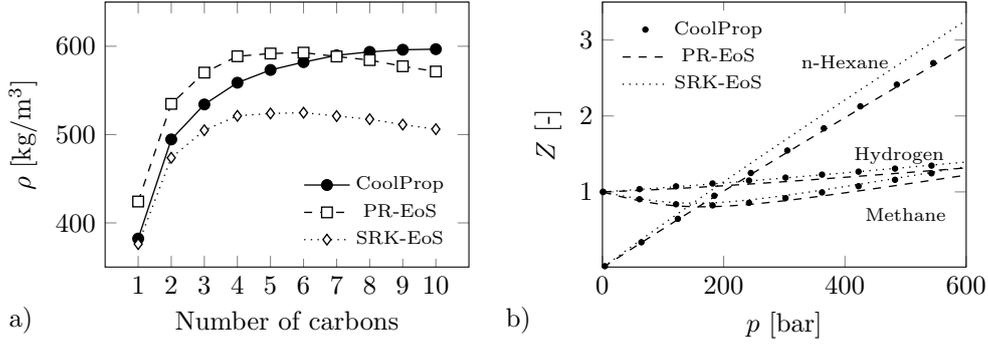

At engine-relevant injection conditions characterized by supercritical pressures and moderate temperatures real-gas effects are a prominent feature~\cite{chehroudi2012a,oschwald2006a} and have to be taken into account within the numerical framework to accurately model the fluid behavior. Commonly accepted in computational fluid dynamics (CFD) due to their efficiency and acceptable accuracy are cubic equation of states (EoS) based on the corresponding states principle, which can be written in the following general, pressure-explicit form~\cite{reid1987}:
\begin{equation}
p = \frac{\mathcal{R} \; T}{v - b} - \frac{a \left(T\right)}{v^2 + ubv + wb^2} = \frac{\mathcal{R} \; T}{v - b} - \frac{a_c \; \alpha\left(T\right)}{v^2 + ubv + wb^2}.
\label{eq:cubicEoS}
\end{equation}
Here, $\mathcal{R}$ is the gas constant, $T$ is the static temperature and $v = 1/\rho$ is the specific volume. The parameters $a(T) = a_c \; \alpha(T)$ and $b$ account for the intermolecular attractive and repulsive forces, respectively, and $u$ and $w$ are model constants. The most popular cubic EoS-models are the ones by Peng and Robinson~\cite{peng1976new} (PR-EoS) and Soave, Redlich and Kwong~\cite{soave1972equilibrium} (SRK-EoS) where $u$ and $w$ are (2,-1) and (1,0), respectively. Due to the different model constants, $a_c$ and $b$ are different among the cubic EoS-models which results in turn in different EoS-specific critical compressibility factors $Z_{c,\textrm{PR}} = 0.307$ for the PR-EoS and $Z_{c,\textrm{SRK}} = 0.333$ for the SRK-EoS. As the critical compressibility factor is a fluid specific parameter, see Tab.~\ref{tab:criticalProperties}, neither the PR-EoS nor the SRK-EoS can predict the $p$,$v$,$T$-behavior of all real fluids accurately and an appropriate EoS-model has to be chosen based on the particular problem investigated. As an example, Fig.~\ref{fig:comparisonAlkanes} shows that the PR-EoS as well as the SRK-EoS can only predict a limited number of alkanes from the homologue series within an acceptable error~\cite{cismondi2005a,kim2012a}. Generally, the density/compressibility factor prediction of the SRK-EoS is better for simple fluids like methane $\textrm{CH}_4$ and hydrogen $\textrm{H}_2$, while the PR-EoS gives superior results for longer alkanes like n-hexane $\textrm{C}_6\textrm{H}_{14}$. This finding can be further emphasized by comparing the isotherms of these different fluids over a wide pressure range, see Fig.~\ref{fig:comparisonAlkanes}~b). For hydrogen the PR-EoS gives a mean deviation from the reference data~\cite{CoolProp} of about -3.81\% and the SRK-EoS of about -0.13\%. In contrast, for n-hexane the mean deviation of the PR-EoS is -1.38\% and for the SRK-EoS it is 10.46\%. In this study, we will investigate three different injectants, namely a hydrogen fuel mixture, compressed natural gas (consisting mostly of methane) and n-hexane. In order to get the best fluid modeling possible, we will use the PR-EoS for n-hexane and the SRK-EoS for the other two fuels containing hydrogen and methane.\\  

\begin{table}[h]
	\centering
	\caption{Parameters of the general cubic equation of state.}
	\vspace{6pt}
	\begin{tabular}{c|cc}
		\toprule
		Parameter  	& PR-EoS~\cite{peng1976new} & SRK-EoS~\cite{soave1972equilibrium} \\
		\midrule
		$u$ 		& 2 	& 1 	\\
		$w$ 		& -1 	& 0 	\\
		$a_c$ 		& $0.45724 \; \frac{\mathcal{R}^2 T_c^2}{p_c}$ & $0.42748 \; \frac{\mathcal{R}^2 T_c^2}{p_c}$ \\
		$\alpha \left( T \right)$ & \multicolumn{2}{c}{$\left[ 1 + \kappa \left( 1 - \sqrt{\frac{T}{T_c}}\right)\right]^2$} \\
		$\kappa$ 	& $0.37464 + 1.54226 \; \omega - 0.26992 \; \omega^2$ & $0.480 + 1.574 \; \omega - 0.176 \; \omega^2$ \\
		$b$ 		& $0.07780 \; \frac{\mathcal{R} T_c}{p_c}$ & $0.08664 \; \frac{\mathcal{R} T_c}{p_c}$ \\
		$Z_c$ 		& $0.307$ & $0.333$ \\
		\bottomrule 
	\end{tabular}
	\label{tab:cubicEoS}
	\vspace{12pt}
\end{table}

For multi-component mixtures we are using the widely-applied concept of an one-fluid mixture in combination with mixing rules~\cite{poling2001} for calculating $a$ and $b$ independent of the chosen EoS:
\begin{equation}
a = \sum_{i}^{N_c}\sum_{j}^{N_c} \xi_i \xi_j a_{ij} \hspace{1cm} \text{and} \hspace{1cm} b = \sum_{i}^{N_c} \xi_i b_i \; .
\label{eq:ab_PR-EoS}
\end{equation}

Here, $\xi_i$ is the mole fraction of species $i$, whereby in the following we denote the overall mole fraction by $\mathbf{z} = \{ z_1 , ... , z_{N_c} \}$ and the liquid and vapor mole fractions by $\mathbf{x} = \{ x_1 , ... , x_{N_c} \}$ and $\mathbf{y} = \{ y_1 , ... , y_{N_c} \}$, respectively. The variables $a_{ij}$ and $b_i$ in Eq.~\eqref{eq:ab_PR-EoS} are calculated using the corresponding state principle and can therefore be evaluated dependent on the EoS, see Tab.~\ref{tab:cubicEoS}. For the calculation of the diagonal elements of $a_{ij}$ the respective critical parameters of the pure components are used, see Tab.~\ref{tab:criticalProperties}. The off-diagonal elements of $a_{ij}$ are estimated using the pseudo-critical combination rules~\cite{reid1987}:
\begin{equation}
\begin{array}{l}
\omega_{ij} = 0.5 \left( \omega_i + \omega_j \right) \; , \;\;\; v_{c,ij} = \frac{1}{8} \left( v_{c,i}^{1/3} + v_{c,j}^{1/3} \right)^3 \; , \;\;\; Z_{c,ij} = 0.5 \left( Z_{c,i} + Z_{c,j} \right) \; , \;\;\;\\
\vspace{0pt}\\
T_{c,ij} = \sqrt{T_{c,i} T_{c,j}} \left(1- k_{ij} \right) \hspace{1cm} \textrm{and} \hspace{1cm} p_{c,ij} = Z_{c,ij} \mathcal{R} T_{c,ij} /v_{c,ij} \; .\\
\vspace{-8pt}
\end{array}
\label{eq:pseudoCritical}
\end{equation}
The binary interaction parameter $k_{ij}$ in Eq.~\eqref{eq:pseudoCritical} is usually used to fit the mixture to available measurement data. For this study we are setting $k_{ij}$ to zero and will discuss later how this assumption compares to the experimental data of the appropriate multi-component mixtures investigated.\\
For the calculation of the caloric properties, like enthalpy and specific heat, the departure function formalism is used, see, e.g., Poling \textit{et~al.}~\cite{poling2001}. The reference condition is determined using the seven-coefficient NASA polynomials proposed by Goos \textit{et~al.}~\cite{NASA}. The viscosity and the thermal conductivity are modeled with the empirical correlation proposed by Chung \textit{et~al.}~\cite{chung1988a}.

\subsection{Phase Separation Modelling}

\begin{figure}[h]
	\centering
	\tikzsetnextfilename{criticalLocus}%
	\input{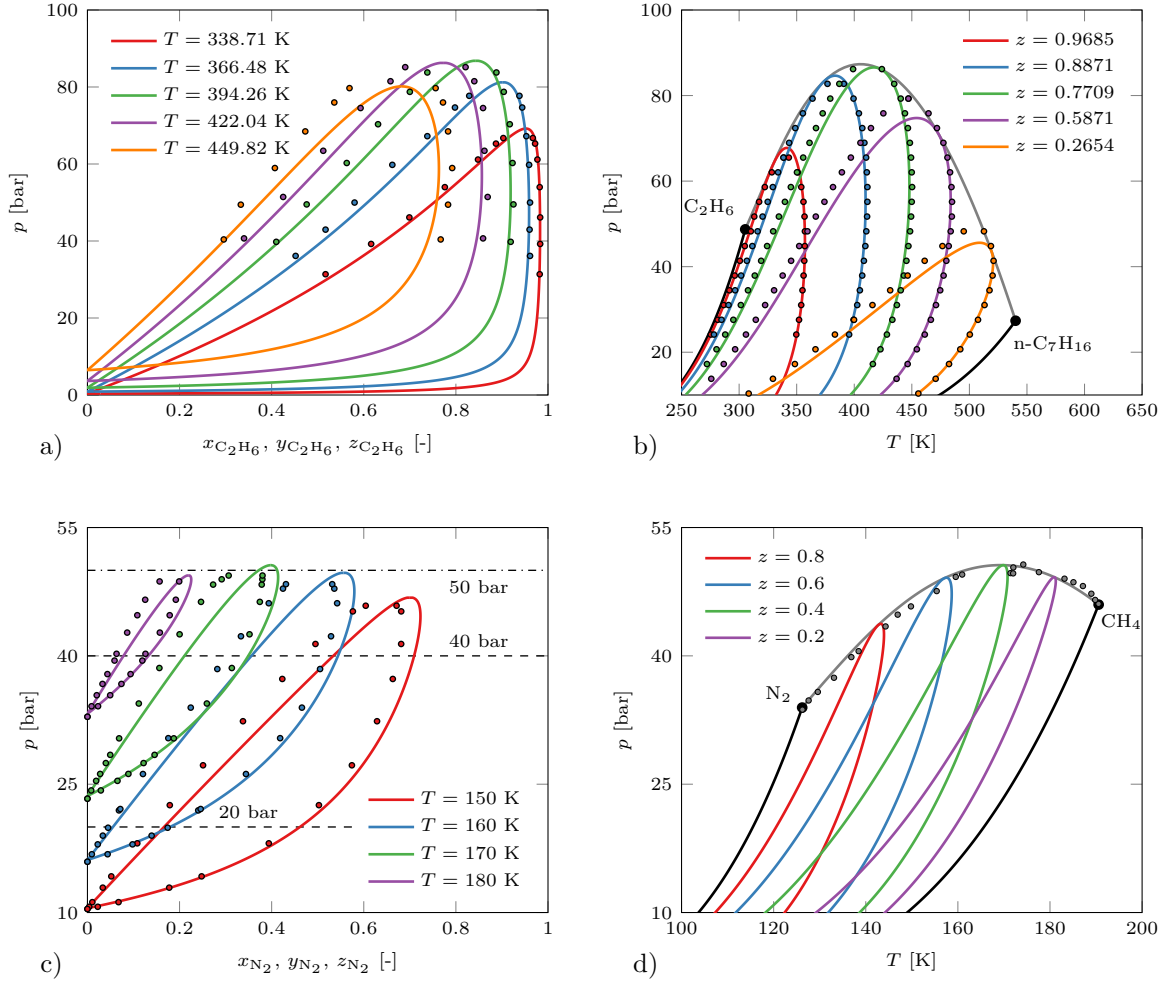}%

	\caption{VLEs for binary mixtures of type I compared to experimental data \cite{kay1938a,mehra1965a,kidnay1975a,chang1967a,bloomer1953a,stryjek1974a,cines1953a} (shown as dots): a) $pxy$-diagram for ethane + n-heptane, b) $pT$-diagram for ethane + n-heptane, c) $pxy$-diagram for nitrogen + methane and d) $pT$-diagram for nitrogen + methane. The black line shows the saturation curve of the pure component and the black dot marks the pure component's critical point. The gray line is the critical locus of the binary mixture calculated based on the approach of Heidemann and Khalil~\cite{heidemann1980a}. The PR-EoS was used to calculate the ethane + n-heptane mixture. The SRK-EoS was used to calculate the nitrogen + methane mixture.}
	\label{fig:criticalLocus}
\end{figure}

Cubic equations of state are generally suitable to describe the complete $p$,$v$,$T$-behavior of single-component as well as multi-component fluids. Therefore, they are also able to predict the phase separation in fluids which might occur when a fluid enters the subcritical regime (temperature and pressure below the critical values). In contrast to single component fluids, which have a distinct, fluid specific critical point, see black dots in Fig.~\ref{fig:criticalLocus}, multi-component mixtures show a critical locus, see gray line in Fig.~\ref{fig:criticalLocus}, reaching over a finite temperature and pressure range. For binary mixtures with type I phase behavior the critical locus is spanning a line from the low volatile component to the high volatile component~\cite{qiu2015a,vanKonynenburg1980a}, see Fig.~\ref{fig:criticalLocus}. The mixture critical temperature is limited by the pure components values while the critical pressure of the mixture often by far exceeds the critical pressure of the pure components~\cite{elliott2012a}. Thus, there is a clear upper border in temperature above which no phase separation will occur, namely the temperature of the high volatile component (Fig.~\ref{fig:criticalLocus}~b): $T_{\textrm{c},\textrm{n-C}_7\textrm{H}_{16}}$ = \SI{469.7}{\kelvin}; Fig.~\ref{fig:criticalLocus}~d): $T_{\textrm{c},\textrm{CH}_{4}}$ = \SI{190.56}{\kelvin}). In terms of critical pressure, Fig.~\ref{fig:criticalLocus} shows that the specific form of the critical locus and therefore the maximum critical pressure is strongly dependent on the pure components forming the mixture. Compared to the critical pressure of the high volatile component, the maximum mixture critical pressure for the binary mixture ethane + n-heptane is approximately a factor of \SI{3.2}{} larger, whereas in the case of the binary mixture nitrogen + methane it is approximately a factor of \SI{1.1}{} only. From Fig.~\ref{fig:criticalLocus} it is also obvious that cubic EoS are able to predict the vapor-liquid equilibria of different binary mixtures within an reasonable accuracy over a wide temperature and pressure range.\\

\begin{table}[h]
	\centering
	\caption{Critical properties ($p_{\textrm{c}}$, $T_{\textrm{c}}$ and $Z_{\textrm{c}}$) and acentric factor $\omega$ of hydrogen $\textrm{H}_2$, methane $\textrm{CH}_4$, ethane C$_2$H$_6$, n-hexane C$_6$H$_{14}$ and nitrogen $\textrm{N}_2$ taken from CoolProp~\cite{CoolProp}.}
	\vspace{6pt}
	\begin{tabular}{c|cccc}
		\toprule
		\multirow{ 2}{*}{Specie}  & $p_{\textrm{c}}$ & $T_{\textrm{c}}$ & $Z_{\textrm{c}}$ & $\omega$  \\
		& [MPa] & [K] & [-] & [-] \\
		\midrule 
		Hydrogen ($\textrm{H}_2$) & 1.2964 & 33.15 & 0.303 & -0.2190 \\
		Methane ($\textrm{CH}_4$) & 4.5992 & 190.56 & 0.286 & 0.0114 \\
		Ethane (C$_2$H$_6$) & 4.8722 & 305.32 & 0.280 & 0.099 \\
		n-Hexane (C$_6$H$_{14}$) & 3.0340 & 507.82 & 0.266 & 0.299 \\
		Nitrogen ($\textrm{N}_2$) & 3.3958 & 126.19 & 0.289 & 0.0372 \\
		\bottomrule 
	\end{tabular}
	\label{tab:criticalProperties}
	\vspace{12pt}
\end{table}

The vapor-liquid equilibria (VLEs) in Fig.~\ref{fig:criticalLocus} are calculated based on the assumption of a thermodynamic equilibrium between the different phases, namely a liquid and a vapor phase. This equilibrium is characterized by the equality of temperature, pressure and chemical potential/specific Gibbs energy among the different phases. As a result, the Gibbs energy in isothermal-isobaric systems under such equilibrium conditions is constant in the total system and is at the global minimum~\cite{luedecke2013a}. In order to realize this minimum, a phase separation takes places if and only if the Gibbs energy can be reduced due to this separation compared to the single phase state or the state with lesser phases, see, e.g., Michelsen~and~Mollerup~\cite{michelsen2007}. As we will only consider a maximum of two phases within this study, a phase separation and therefore a reduction in Gibbs energy will only be executed from the single-phase state to a two-phase state. To take into account this possible phase separation process within the present study, we are using a vapor-liquid equilibrium (VLE) model, see, e.g., Matheis~and~Hickel~\cite{matheis2016a} and Traxinger \textit{et~al.}~\cite{traxinger2017a}. In this model, the stability check is done with the help of the tangent plane distance (TPD) method of Michelsen~\cite{michelsen1982a}
\begin{equation}
TPD \left( \mathbf{w} \right) = \sum_{i} w_i \left[ \ln w_i  + \ln \varphi_i \left(\mathbf{w}\right) - \ln z_i  - \ln \varphi_i \left(\mathbf{z}\right) \right] \; ,
\label{eq:TPD}
\end{equation}
where the Gibbs energy of a trial phase composition $\mathbf{w} = \{ w_1 , ... , w_{N_c} \}$ is compared to the Gibbs energy of the respective feed mixture $\mathbf{z}$. The Gibbs energy is expressed based on the fugacity coefficient of component $i$ $\varphi_i$ which is defined as $f_i^l/ \left( x_i p \right)$ for the liquid and as $f_i^v/ \left( y_i p \right)$ for the vapor phase, respectively. The fugacity of component $i$ in the respective phase is denoted by $f_i$. The natural logarithm of $\varphi_i$ can be calculated directly from the applied cubic EoS:
\begin{equation}
\textrm{ln} \left( \varphi_i \right) = \frac{\partial \left(F - F^{\textrm{ig}}\right)/\left(R T\right)}{\partial n_i}\at[\Bigg]{T,V} - \textrm{ln} \left( Z \right) \; .
\end{equation}
Here, $\frac{\partial \left(F - F^{\textrm{ig}}\right)/\left(R T\right)}{\partial n_i}\at[\big]{T,V}$ is the partial derivative of the departure function of the normalized Free Helmholtz energy $F$ with respect to the mole number of the $i$-th component $n_i$. This partial derivative reads
\begin{equation*}
-\textrm{ln} \left( 1 - b \rho \right) + \frac{b_i}{b} \left( Z - 1 \right) - \frac{a}{\sqrt{8} b R T} \left( \frac{2 \sum_{j} x_j a_{ij}}{a} - \frac{b_i}{b} \right) \textrm{ln} \left[ \frac{1+b \rho \left( 1 + \sqrt{2} \right)}{1+b \rho \left( 1 - \sqrt{2} \right)} \right]
\end{equation*}
for the PR-EoS and 
\begin{equation*}
-\textrm{ln} \left( 1 - b \rho \right) + \frac{b_i}{b} \left( Z - 1 \right) - \frac{a}{b R T} \left( \frac{2 \sum_{j} x_j a_{ij}}{a} - \frac{b_i}{b} \right) \textrm{ln} \left( 1 + \frac{b}{v} \right)
\end{equation*}
for the SRK-EoS. If the TPD-analysis of Eq.~\eqref{eq:TPD} leads to a negative value for any of the trial phase compositions $\mathbf{w}$, the mixture is unstable. In this investigation, a separation in two phases, namely a vapor ($v$) and a liquid ($l$) phase, is done yielding a decrease in Gibbs energy. It is important to notice that this phase separation is assumed to occur instantaneously, so no non-equilibrium effects are taken into account, which might be present especially at high Mach numbers as it is for example the case in last stages of low-pressure steam turbines~\cite{gyarmathy1962a}. After the detection of an unstable mixture an iso-energetic and isobaric flash is solved. Due to the fact that we are using an enthalpy based energy conservation equation the flash can be considered a $hpn$-flash which is constructed as a nested loop whereby the outer loop is updating the temperature in order to meet the energy-criterion and in the inner loop a $Tpn$-flash for the appropriate temperature is solved. The solution of the $Tpn$-flash is characterized by the equality of the fugacities of each component in the considered phases~\cite{elliott2012a}:
\begin{equation}
f_i^l (p,T,\mathbf{x}) = f_i^v (p,T,\mathbf{y}) \; .
\end{equation}
For solving the TPD-analysis and the $Tpn$-flash we strongly followed the suggestions of Michelsen and Mollerup~\cite{michelsen2007} and implemented three different methods, a successive substitution method (SSM), a dominant eigenvalue method (DEM) and Newton's method (NM). Usually, we are starting with the successive substitution for solving both problems and after a certain number of evaluation steps we are switch to the higher order methods (DEM or NM) to speed-up the iteration process. This is also important for near-critical calculations as the SSM might not converge or only with a very large number of iterations~\cite{michelsen2007}.

\subsection{Compressibility and Speed of Sound in Two-Phase Mixtures}
\label{subsec:Psi}

\subsubsection{Methodology}

Compressibility effects play an important role in real-gas flows and flows with high Mach numbers. Basically, three different definitions are used to define compressibilities in thermodynamics, the isothermal compressibility $\psi_T$, the isentropic compressibility $\psi_s$ and the isenthalpic compressibility $\psi_h$, see Tab.~\ref{tab:thermoParas}. These three compressibilities can be related to each other by the Gr\"uneisen parameter~\cite{arp1984a} see Tab.~\ref{tab:thermoParas},
\begin{equation}
\phi = v \frac{\partial p}{\partial e}\at[\Bigg]{v} = \frac{v}{c_v} \frac{\partial p}{\partial T}\at[\Bigg]{v} = \frac{\rho}{T} \frac{\partial T}{\partial \rho}\at[\Bigg]{s}
\label{eq:Grueneisen1}
\end{equation}
which can also be expressed by different general thermodynamic response functions \cite{mausbach2016a} as:
\begin{equation}
\phi = \frac{v \alpha_p}{c_v \psi_T} = \frac{\alpha_p a_s^2}{c_p} \; .
\label{eq:Grueneisen2}
\end{equation}
In Equations~\eqref{eq:Grueneisen1} and \eqref{eq:Grueneisen2} $e$ is the internal energy, $c_v$ is the isochoric heat capacity and $c_p$ is the isobaric heat capacity. In fluids without hydrogen bondings, as it is the case for this study, the Gr\"unstein parameter is always greater than zero because this type of fluids are having no density anomalies like, e.g., water has~\cite{mausbach2016a}. As a result, one can deduce from the equations in Tab.~\ref{tab:thermoParas} that
\begin{equation}
\psi_T > \psi_h > \psi_s \; .
\label{eq:relationPsi}
\end{equation}
It is very important to distinguish between these three compressibilities and to choose the suitable one depending on the application and numerical framework. In our case we are using the isenthalpic compressibility in the pressure equation as the energy conservation equation is solved in terms of an enthalpy equation. The second compressibility being highly important within this study is the isentropic compressibility because the speed of sound $a_s$, which is an important ingredient of the KT scheme, is a function of the inverse of the isentropic compressibility $\psi_s$ and the density $\rho$:
\begin{equation}
a_s = \sqrt{\frac{\partial p}{\partial \rho}\at[\Bigg]{s,z_i}} = \sqrt{-v^2 \; \frac{\partial p}{\partial v}\at[\Bigg]{s,z_i}} = \frac{1}{\sqrt{\rho  \; \psi_s}} \; .
\label{eq:as}
\end{equation}\\

\begin{table}[t]
	\vspace{0pt}
	\centering
	\caption{Definition of thermodynamic parameters.}
	\vspace{6pt}
	\begin{tabular}{l|cc}
		\toprule
		Name & Symbol & Equation \\
		\midrule 
		Isothermal compressibility & $\psi_T$ & $-\frac{1}{v}\frac{\partial v}{\partial p}\at[\big]{T} = \psi_s\left(1 + T \alpha_p \phi \right)$  \\
		Isentropic compressibility & $\psi_s$ & $-\frac{1}{v}\frac{\partial v}{\partial p}\at[\big]{s}$  \\
		Isenthalpic compressibility & $\psi_h$ & $-\frac{1}{v}\frac{\partial v}{\partial p}\at[\big]{h} = \psi_s \left( 1 + \phi \right)$  \\
		Thermal expansivity & $\alpha_p$ & $\frac{1}{v}\frac{\partial v}{\partial T}\at[\big]{p}$  \\
		\bottomrule 
	\end{tabular}
	\label{tab:thermoParas}
	\vspace{12pt}
\end{table}

In single phase systems the calculation of the speed of sound and the isenthalpic compressibility is almost straightforward, although it implies the calculation of some partial derivatives, which can be solved analytically for the cubic EoS used in this investigation, see, e.g., M\"uller \textit{et~al.}~\cite{muller2016b} for the PR-EoS. In multi-phase systems this is not true anymore and the evaluation gets way more complicated. This is also true for measurements and therefore many researchers focus on the numerical investigation of the speed of sound in multi-component, multi-phase systems, e.g., Picard~and~Bishnoi~\cite{picard1987a}, Firoozabadi~and~Pan~\cite{firoozabadi2000a}, Nichita~\textit{et~al.}~\cite{nichita2010a} and Castier~\cite{castier2011a}. A simplified and very popular method for calculating the speed of sound in two-phase systems is the correlation of Wood~\cite{wood1930a}. This correlation is based on a volume-weighted approach of the bulk modulus respectively its inverse, the isentropic compressibility
\begin{equation}
\psi_s = \left(1 - \beta_v\right) \, \psi_{s,\textrm{l}}\left(\mathbf{x},p,T\right) + \beta_v \, \psi_{s,\textrm{v}}\left(\mathbf{y},p,T\right) = \frac{\left(1 - \beta_v\right)}{\rho_\textrm{l} \, a_{s,\textrm{l}}^2} + \frac{\beta_v}{\rho_\textrm{g} \, a_{s,\textrm{g}}^2}
\label{eq:psiSWood}
\end{equation}
where $\beta_v$ denotes the vapor volume fraction. Applying Eq.~\eqref{eq:psiSWood} to Eq.~\eqref{eq:as} yields Wood's correlation for the speed of sound in a two-phase mixture:
\begin{equation}
a_s = \left[ \rho \left( \frac{\left(1 - \beta_v\right)}{\rho_\textrm{l} \, a_{s,\textrm{l}}^2} + \frac{\beta_v}{\rho_\textrm{g} \, a_{s,\textrm{g}}^2} \right) \right]^{\left(-1/2\right)} \; .
\label{eq:asWood}
\end{equation}
This correlation is suitable to capture the basic trends of the variation of the speed of sound in the two-phase region but misses the abrupt/discontinues changes at the phase boundaries, see Nichita \textit{et~al.}~\cite{nichita2010a} for a more thorough discussion. Different approaches can be found in literature to numerically/analytically calculate the thermodynamic speed of sound in two-phase mixtures~\cite{picard1987a,firoozabadi2000a,nichita2010a,castier2011a}. In this study we follow the approach devised by Nichita \textit{et~al.}~\cite{nichita2010a} as it is a general and efficient numerical scheme~\cite{castier2011a}. Here we are able to use the already implemented $Tpn$-flash as a basis to calculate the thermodynamic speed of sound.\\

In a single-phase fluid the isothermal and the isentropic compressibility are related by the Gr\"uneisen parameter, see Tab.~\ref{tab:thermoParas}, but can also be related by the ratio of the specific heat at constant pressure to constant volume $\kappa$:
\begin{equation}
\psi_s = \frac{\psi_T}{\kappa} \; .
\label{eq:psiSpsiT}
\end{equation}
By applying some basic mathematical and thermodynamical relations Eq.~\eqref{eq:psiSpsiT} can be rewritten as~\cite{nichita2010a}
\begin{equation}
\psi_s = \psi_T - \frac{\alpha_p^2 \, T \, v}{c_p} \; ,
\label{eq:psisFinal}
\end{equation}
where $\alpha_p$ is the thermal expansivity, see Tab.~\ref{tab:thermoParas}. This basic derivation holds also for two-phase systems if one takes into account the phase split when evaluating the different properties and partial derivatives, for more details see Nichita \textit{et~al.}~\cite{nichita2010a}. From the Eq.~\eqref{eq:psisFinal} and Tab.~\ref{tab:thermoParas} it is obvious that Eq.~\eqref{eq:psisFinal} only contains partial derivatives at constant temperature and pressure, respectively. Therefore, we followed the suggestion of Nichita \textit{et~al.}~\cite{nichita2010a} and calculate this partially derivatives numerically by applying $Tpn$-flashes at ($p$,$T\pm\epsilon$) and ($p\pm\epsilon$,$T$). These flashes converge very fast within 1-2 Newton iterations, since an excellent initial guess is available from the $hpn$-flash performed during every iteration/time step (if the mixture was characterized unstable by the TPD-analysis).\\

For calculating the isenthalpic compressibility $\psi_h$ we follow the same idea and rewrite the basic definition in such a way that it only contains partial derivatives with respect to temperature and pressure. This can be done by using the chain rules
\begin{equation}
\frac{\partial v}{\partial p}\at[\Bigg]{h} \, \frac{\partial p}{\partial h}\at[\Bigg]{v} \, \frac{\partial h}{\partial v}\at[\Bigg]{p} = -1
\label{eq:chainRulevph}
\end{equation}
and
\begin{equation}
\frac{\partial T}{\partial p}\at[\Bigg]{v} \, \frac{\partial p}{\partial v}\at[\Bigg]{T} \, \frac{\partial v}{\partial T}\at[\Bigg]{p} = -1 \; ,
\label{eq:chainRuleTpv}
\end{equation}
the total derivative of the enthalpy with respect to temperature and pressure
\begin{equation}
\textrm{d} h = \frac{\partial h}{\partial T}\at[\Bigg]{p} \, \textrm{d} T + \frac{\partial h}{\partial p}\at[\Bigg]{T} \, \textrm{d} p \; ,
\label{eq:totalDerh}
\end{equation}
the Gibbs equation
\begin{equation}
\textrm{d} h = T \textrm{d} s + v \textrm{d} p
\label{eq:Gibbs}
\end{equation}
as well as the Maxwell identity
\begin{equation}
\frac{\partial s}{\partial p}\at[\Bigg]{T} = - \frac{\partial v}{\partial T}\at[\Bigg]{p} \; .
\label{eq:Maxwell}
\end{equation}
Applying Eqs.~\eqref{eq:chainRulevph}-\eqref{eq:Maxwell} to the definition of the isenthalpic compressibility, see Tab.~\ref{tab:thermoParas}, yields
\begin{equation}
\psi_h = \frac{1}{v} \frac{-c_p \frac{\partial v}{\partial p}\at[\big]{T}/\frac{\partial v}{\partial T}\at[\big]{p} + v - T \frac{\partial v}{\partial T}\at[\big]{p}}{c_p / \frac{\partial v}{\partial T}\at[\big]{p}}
\end{equation}
and therefore an equation containing only partial derivatives with respect to temperature and pressure. As a result, the speed of sound $a_s$ and the isenthalpic compressibility $\psi_h$ in a multi-component, two-phase system can be calculated numerically by solving the $Tpn$-flashes at ($p$,$T\pm\epsilon$) and ($p\pm\epsilon$,$T$) after the $hpn$-flash-problem has been solved successfully. This makes it numerically efficient compared to iso-energy, iso-volume or iso-entropy flashes because those flashes always require a nested loop or a further objective function in addition to the equality of fugacities in the $Tpn$-flash.\\

\subsubsection{Validation}

\begin{figure}[t]
	\centering
	\tikzsetnextfilename{prudhoeBay}%

\begin{tikzpicture}
	\begin{groupplot}[group style = {group name = plots,group size = 3 by 2, horizontal sep = 50pt, vertical sep = 50pt}, width = 0.33\textwidth, height = 0.33\textwidth]
	\nextgroupplot
	[
	xmin=140,
	xmax=260,
	xtick = {140,160,180,200,220,240,260},
	xlabel = $T\text{ [K]}$,
	ymin=0.1,
	ymax=90,
	ylabel = $p\text{ [bar]}$,
	ytick = {0,10,20,30,40,50,60,70,80,90},
	y filter/.code={\pgfmathparse{#1*1e-5}\pgfmathresult},
	xticklabel style = {font=\scriptsize,yshift=0.5ex},
	yticklabel style = {font=\scriptsize,xshift=0.5ex},
	label style = {font=\scriptsize,xshift=0.5ex}
	]
	 	\addplot [variable=\T,samples=2,domain=140:260,line width=1.0pt,color=color1] ({T},{10e5});\label{plot:10bar}
	 	\addplot [variable=\T,samples=2,domain=140:260,line width=1.0pt,color=color2] ({T},{30e5});\label{plot:30bar}
	 	\addplot [variable=\T,samples=2,domain=140:260,line width=1.0pt,color=color3] ({T},{50e5});\label{plot:50bar}
	 	\addplot [variable=\T,samples=2,domain=140:260,line width=1.0pt,color=color4] ({T},{70e5});\label{plot:70bar}
		\addplot [line width=1.0pt,forget plot] table [x=T, y=p]{./data/prudhoeBay/pT_Diagram_PrudhoeBay.dat};
		\addplot [line width=1.0pt,forget plot,mark=o,draw=none] table [x=T_c, y=p_c]{./data/prudhoeBay/pc_Tc_Diagram_PrudhoeBay.dat};
		\node at (axis cs: 226.3,74.8) [anchor = east,font=\scriptsize] {CP };
		\coordinate (top) at (rel axis cs:0,1);
	\nextgroupplot
	[
    xmin=130,
    xmax=270,
    xtick = {130,150,170,190,210,230,250,270},
    xlabel = $T\text{ [K]}$,
    ymin=1,
    ymax=1200,
    ylabel = $a_s\text{ [m/s]}$,
    ytick = {0,200,400,600,800,1000,1200,1400},
   	xticklabel style = {font=\scriptsize,yshift=0.5ex},
   	yticklabel style = {font=\scriptsize,xshift=0.5ex},
   	label style = {font=\scriptsize,xshift=0.5ex}
    ]
	    \addplot [color=color1,line width=1.0pt,forget plot] table [x=T, y=as]{./data/prudhoeBay/as_PrudhoeBay_p_10bar.dat};
	    \addplot [color=color1,line width=1.0pt,forget plot,dashed] table [x=T, y=asWood]{./data/prudhoeBay/as_PrudhoeBay_p_10bar.dat};
 	    \addplot [color=color2,line width=1.0pt,forget plot] table [x=T, y=as]{./data/prudhoeBay/as_PrudhoeBay_p_30bar.dat};
 	    \addplot [color=color2,line width=1.0pt,forget plot,dashed] table [x=T, y=asWood]{./data/prudhoeBay/as_PrudhoeBay_p_30bar.dat};
   	    \addplot [color=color3,line width=1.0pt,forget plot] table [x=T, y=as]{./data/prudhoeBay/as_PrudhoeBay_p_50bar.dat};
   	    \addplot [color=color3,line width=1.0pt,forget plot,dashed] table [x=T, y=asWood]{./data/prudhoeBay/as_PrudhoeBay_p_50bar.dat};
   	    \addplot [color=color4,line width=1.0pt,forget plot] table [x=T, y=as]{./data/prudhoeBay/as_PrudhoeBay_p_70bar.dat};
   	    \addplot [color=color4,line width=1.0pt,forget plot,dashed] table [x=T, y=asWood]{./data/prudhoeBay/as_PrudhoeBay_p_70bar.dat};
   	    \addplot [variable=\T,samples=2,domain=0:1,line width=1.0pt] ({T},{-1*T});\label{plot:thermo};
   	    \addplot [variable=\T,samples=2,domain=0:1,dashed,line width=1.0pt] ({T},{-1*T});\label{plot:Wood};
	\nextgroupplot
	[
	xmin=130,
	xmax=270,
	xlabel = $T\text{ [K]}$,
	xtick = {130,150,170,190,210,230,250,270},
	ymin=0.01,
	ymax=0.4,
	ylabel = $\psi_s\text{ [1/bar]}$,
	y filter/.code={\pgfmathparse{#1*1e5}\pgfmathresult},
	legend style={legend pos=north east,font=\scriptsize,draw=none},
	legend cell align={left},
	xticklabel style = {font=\scriptsize,yshift=0.5ex},
	yticklabel style = {font=\scriptsize,xshift=0.5ex},
	label style = {font=\scriptsize,xshift=0.5ex}
	]
		\addplot [color=color1,line width=1.0pt,forget plot] table [x=T, y=betas]{./data/prudhoeBay/as_PrudhoeBay_p_10bar.dat};
		\addplot [color=color1,line width=1.0pt,forget plot,dashed] table [x=T, y=betasWood]{./data/prudhoeBay/as_PrudhoeBay_p_10bar.dat};
		\addplot [color=color2,line width=1.0pt,forget plot] table [x=T, y=betas]{./data/prudhoeBay/as_PrudhoeBay_p_30bar.dat};
		\addplot [color=color2,line width=1.0pt,forget plot,dashed] table [x=T, y=betasWood]{./data/prudhoeBay/as_PrudhoeBay_p_30bar.dat};
		\addplot [color=color3,line width=1.0pt,forget plot] table [x=T, y=betas]{./data/prudhoeBay/as_PrudhoeBay_p_50bar.dat};
		\addplot [color=color3,line width=1.0pt,forget plot,dashed] table [x=T, y=betasWood]{./data/prudhoeBay/as_PrudhoeBay_p_50bar.dat};
		\addplot [color=color4,line width=1.0pt,forget plot] table [x=T, y=betas]{./data/prudhoeBay/as_PrudhoeBay_p_70bar.dat};
		\addplot [color=color4,line width=1.0pt,forget plot,dashed] table [x=T, y=betasWood]{./data/prudhoeBay/as_PrudhoeBay_p_70bar.dat};
	\nextgroupplot
	[
	xmin=130,
	xmax=270,
	xlabel = $T\text{ [K]}$,
	xtick = {130,150,170,190,210,230,250,270},
	ymin=0.01,
	ymax=0.4,
	ylabel = $\psi_h\text{ [1/bar]}$,
	y filter/.code={\pgfmathparse{#1*1e5}\pgfmathresult},
	legend style={legend pos=north east,font=\scriptsize,draw=none},
	legend cell align={left},
	xticklabel style = {font=\scriptsize,yshift=0.5ex},
	yticklabel style = {font=\scriptsize,xshift=0.5ex},
	label style = {font=\scriptsize,xshift=0.5ex}
	]
		\addplot [color=color1,line width=1.0pt,forget plot] table [x=T, y=betah]{./data/prudhoeBay/as_PrudhoeBay_p_10bar.dat};
		\addplot [color=color2,line width=1.0pt,forget plot] table [x=T, y=betah]{./data/prudhoeBay/as_PrudhoeBay_p_30bar.dat};
		\addplot [color=color3,line width=1.0pt,forget plot] table [x=T, y=betah]{./data/prudhoeBay/as_PrudhoeBay_p_50bar.dat};
		\addplot [color=color4,line width=1.0pt,forget plot] table [x=T, y=betah]{./data/prudhoeBay/as_PrudhoeBay_p_70bar.dat};
	\nextgroupplot
	[
	xmin=130,
	xmax=270,
	xlabel = $T\text{ [K]}$,
	xtick = {130,150,170,190,210,230,250,270},
	ymin=0.01,
	ymax=0.8,
	ylabel = $\psi_T\text{ [1/bar]}$,
	y filter/.code={\pgfmathparse{#1*1e5}\pgfmathresult},
	legend style={legend pos=north east,font=\scriptsize,draw=none},
	legend cell align={left},
	xticklabel style = {font=\scriptsize,yshift=0.5ex},
	yticklabel style = {font=\scriptsize,xshift=0.5ex},
	label style = {font=\scriptsize,xshift=0.5ex}
	]
		\addplot [color=color1,line width=1.0pt,forget plot] table [x=T, y=betaT]{./data/prudhoeBay/as_PrudhoeBay_p_10bar.dat};
		\addplot [color=color2,line width=1.0pt,forget plot] table [x=T, y=betaT]{./data/prudhoeBay/as_PrudhoeBay_p_30bar.dat};
		\addplot [color=color3,line width=1.0pt,forget plot] table [x=T, y=betaT]{./data/prudhoeBay/as_PrudhoeBay_p_50bar.dat};
		\addplot [color=color4,line width=1.0pt,forget plot] table [x=T, y=betaT]{./data/prudhoeBay/as_PrudhoeBay_p_70bar.dat};
	\nextgroupplot
	[
	xmin=130,
	xmax=270,
	xlabel = $T\text{ [K]}$,
	xtick = {130,150,170,190,210,230,250,270},
	ymin=0.01,
	ymax=1.5,
	ylabel = $\phi\text{ [-]}$,
	legend style={legend pos=north east,font=\scriptsize,draw=none},
	legend cell align={left},
	xticklabel style = {font=\scriptsize,yshift=0.5ex},
	yticklabel style = {font=\scriptsize,xshift=0.5ex},
	label style = {font=\scriptsize,xshift=0.5ex}
	]
		\addplot [color=color1,line width=1.0pt,forget plot] table [x=T, y=Grueneisen]{./data/prudhoeBay/as_PrudhoeBay_p_10bar.dat};
		\addplot [color=color2,line width=1.0pt,forget plot] table [x=T, y=Grueneisen]{./data/prudhoeBay/as_PrudhoeBay_p_30bar.dat};
		\addplot [color=color3,line width=1.0pt,forget plot] table [x=T, y=Grueneisen]{./data/prudhoeBay/as_PrudhoeBay_p_50bar.dat};
		\addplot [color=color4,line width=1.0pt,forget plot] table [x=T, y=Grueneisen]{./data/prudhoeBay/as_PrudhoeBay_p_70bar.dat};
		\coordinate (bot) at (rel axis cs:1,0);
	\end{groupplot}	
    \node[below = 0.3cm of plots c1r1.south,xshift=-2.0cm] {a)};
   	\node[below = 0.3cm of plots c2r1.south,xshift=-2.0cm] {b)};
   	\node[below = 0.3cm of plots c3r1.south,xshift=-2.0cm] {c)};
    \node[below = 0.3cm of plots c1r2.south,xshift=-2.0cm] {d)};
	\node[below = 0.3cm of plots c2r2.south,xshift=-2.0cm] {e)};
 	\node[below = 0.3cm of plots c3r2.south,xshift=-2.0cm] {f)};
   	\path (top|-current bounding box.north)--
	      coordinate(legendpos)
	      (bot|-current bounding box.north);
   	\node[
	anchor=south,
	inner sep=0.2em,
	font=\scriptsize,
	draw
	]at([xshift=0ex,yshift=2ex]legendpos)
	{
	\ref{plot:10bar} 10 bar \hspace*{12pt} \ref{plot:30bar} 30 bar \hspace*{12pt} \ref{plot:50bar} 50 bar \hspace*{12pt} \ref{plot:70bar} 70 bar \hspace*{12pt} \rule[-0.4ex]{0.2ex}{1.2em} \hspace*{12pt} \ref{plot:thermo} Thermodynamics (PR-EoS) \hspace*{12pt} \ref{plot:Wood} Wood
	};
\end{tikzpicture}%

	\caption{Investigation of the 14 component Prudhoe Bay mixture \cite{picard1987a} at four different pressure levels (10 bar, 30 bar, 50 bar and 70 bar): a) Vapor-liquid equilibrium, b) Speed of sound (solid: thermodynamic calculation, dashed: Wood's correlation~\cite{wood1930a}), c) Isentropic compressibility, d) Isenthalpic compressibility, e) Isothermal compressibility and f) Gr\"uneisen parameter. For all diagrams the PR-EoS was used.}
	\label{fig:prudhoeBayAsVLE}
\end{figure}
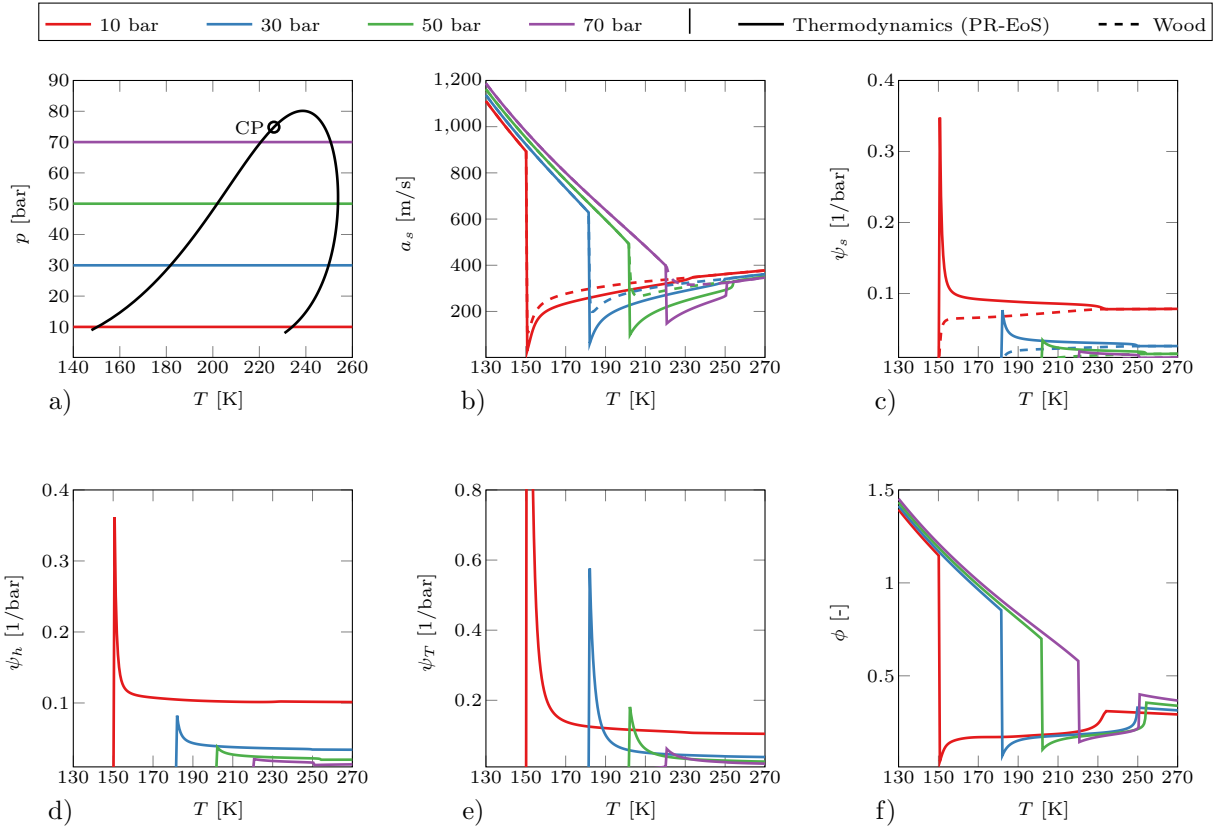

Different research groups~\cite{picard1987a,nichita2010a,castier2011a} have used the Prudhoe Bay gas mixture to study the variation of the speed of sound inside the two-phase region by means of numerics and by applying the PR-EoS to model the real-gas behavior. We will also follow their example in this study but will go beyond this and also discuss additional thermodynamic characteristics like the isenthalpic compressibility and the Gr\"uneisen parameter, which relates the different compressibilities to each other. The Prudhoe Bay mixture contains 14 components with the following mole fractions listed in parentheses~\cite{picard1987a}: methane (83.3310), ethane (9.6155), propane (3.5998), iso-butane (0.3417), n-butane (0.4585), iso-pentane (0.0403), n-pentane (0.0342), n-hexane (0.0046), n-heptane (0.0003), n-octane (0.0001), toluene (0.0002), nitrogen (1.4992), oxygen (0.0008) and carbon-dioxide (1.0738). In Fig.~\ref{fig:prudhoeBayAsVLE}~a) the VLE of this 14 component mixture is shown together with the four pressure levels used for the more thorough investigation of this mixture. The shape of the VLE is in very good agreement with the results of Picard~and~Bishnoi~\cite{picard1987a}. However, the critical point is estimated at a different location (Picard~and~Bishnoi~\cite{picard1987a}: 206 K, 5.5 MPa; present study: 226.3 K, 7.48 MPa). 
The speed of sound shown in Fig.~\ref{fig:prudhoeBayAsVLE}~b) is in very good agreement with the results of other research groups \cite{picard1987a,nichita2010a,castier2011a}. At the bubble-point line (low temperatures) a drastic, discontinuous jump is observed, which reduces with increasing pressure. For the lowest pressure (\SI{10}{\bar}) the speed of sound is almost dropping by a factor of 30 as the fluid crosses the coexistence-line and first vapor bubbles are forming. Afterwards, as the fluid is heating up and the vapor fraction is steadily rising, the speed of sound is increasing continuously. A small discontinuity is also observed at the dew-point line (large temperatures). This jump gets larger with increasing pressure which is contrary to the behavior of $a_s$ at the bubble-point line. Outside of the two-phase region the fluid behaves liquid-like to the left of the VLE resulting in an increase of $a_s$ with reducing temperature. On the right side of the VLE the opposite is true as the fluid behaves gas-like. The comparison of Wood's correlation~\cite{wood1930a} with the thermodynamic speed of sound in Fig.~\ref{fig:prudhoeBayAsVLE}~b) underlines the statement that the correlation is able to capture the basic trend within the two-phase region but misses the significant drops of the speed of sound at the phase boundaries, compare Nichita \textit{et~al.}~\cite{nichita2010a} for similar findings. This observation holds also for the isentropic compressibility as it behaves indirectly proportional to the speed of sound, see Fig.~\ref{fig:prudhoeBayAsVLE}~c) and Eq.~\eqref{eq:as}.\\
In order to do a further evaluation of the numerical approach described above, we also show the different compressibilities ($\psi_s$, $\psi{_h}$ and $\psi_T$) in Fig.~\ref{fig:prudhoeBayAsVLE}. The comparison of the plots shows that the basic relation between the three compressibilities $\psi_T > \psi_h > \psi_s$, see Eq.\eqref{eq:relationPsi}, holds. Furthermore, Figs.~\ref{fig:prudhoeBayAsVLE}~c)-e) show that the isenthalpic and the isentropic compressibility are very similar within the VLE. In contrast, the isothermal compressibility is much larger than the other two ones (note the scaling of the $y$-axis by a factor of two). The Gr\"uneisen parameter in Fig~\ref{fig:prudhoeBayAsVLE}~f) shows a very similar pattern compared to the speed of sound outside as well as inside of the VLE. The identical pattern of $a_s$ and $\phi$ was also reported by Arp~\textit{et~al.}~\cite{arp1984a}, who found this kind of behavior for single component fluids. As an overall result of this discussion we conclude that the thermodynamic framework is able to calculate both the speed of sound as well as the isenthalpic compressibility inside the two-phase region.

\section{Validation of the VLE-model in OpenFOAM}
\label{sec:ILASS}

Recently, experiments and numerical simulations focusing on the process of phase separation due to multi-component mixing at high-pressure conditions were conducted by Traxinger \textit{et al.}~\cite{traxinger2017a}. In the experiments n-hexane was injected into a pressurized chamber with pure nitrogen ($p_{ch}=\SI{50}{\bar}$, $T_{ch}=\SI{293}{\kelvin}$) at rest. The injected fluid was heated to three different temperatures such that the test cases cover regimes in which thermodynamic non-idealities, namely phase-separation induced by multi-component mixing at supercritical pressure with respect to the pure components values, are significant. To capture the flow structure as well as the phase separation in the experiments, shadowgraphy and elastic light scattering were performed simultaneously. For additional information about the experimental setup see Lamanna \textit{et~al.}~\cite{lamanna2012a} and Baab \textit{et~al.}~\cite{baab2014a}. In addition to the experiments, large eddy simulations were carried out applying a VLE-model. The results were in qualitatively very good agreement with the experimental findings in terms of phase separation and overall flow structure. In order to validate the thermodynamic modeling coupled with the hybrid solver, numerical simulations with the same boundary conditions were performed, see Tab.~\ref{tab:validationBC}. As the boundary conditions for the experimental setup were only given as static inlet properties, the corresponding total plenum conditions were calculated under the constant isentropic and isenthalpic nozzle flow assumptions~\cite{banholzer2017}:
\begin{equation}
s_0 = s_0\left(p_0,T_0\right) = s\left(p,T\right) \qquad h_0 = h_0\left(p_0,T_0\right) = h\left(p,T\right) + \frac{1}{2} u^2.
\end{equation}

For the three test cases T600, T560 and T480, named after the total temperature $T_{t,inlet}$ of the injected n-hexane, the resulting total pressures are \SI{56.40}{\bar}, \SI{56.57}{\bar} and \SI{54.56}{\bar}. Case T600 with the highest injection temperature of \SI{600}{\kelvin} is likely to be a dense-gas jet and is therefore showing no phase separation effects. For a decreased temperature of \SI{560}{\kelvin} minor two-phase effects can be expected while a temperature decrease to \SI{480}{\kelvin} will show strong two-phase effects in form of a spray-like jet~\cite{traxinger2017a}.
\begin{table}[h]
	\vspace{6pt}
	\centering
	\caption{Overview of the numerical boundary conditions used in the validation cases.}
	\vspace{6pt}
	\begin{tabular}{c|c|cccccc}
		\toprule
		\multirow{ 2}{*}{Case}	& \multirow{ 2}{*}{EoS} 	& $p_{t,fuel}$ & $p_{fuel}$ & $T_{t,fuel}$ & $T_{fuel}$ & $p_{ch}$ & $T_{ch}$ \\
								&							& [bar] 	   & [bar] 	    & [K] 		   & [K] 	    & [bar]    & [K] \\
		\midrule
		T600 					& \multirow{ 3}{*}{PR}		& 56.40 	   & 50.0 		& 600.0		   & 595.0 		& 50.0 	   & 293.0\\
		T560 					&							& 56.57 	   & 50.0 		& 560.0		   & 554.8 		& 50.0 	   & 293.0\\
		T480 					&							& 54.56 	   & 50.0 		& 480.0		   & 479.3 		& 50.0 	   & 293.0\\
		\bottomrule 
	\end{tabular}
	\label{tab:validationBC}
	\vspace{12pt}
\end{table}

\begin{figure}[h]
	\centering
	\tikzsetnextfilename{testcase_schematic}%

	\begin{tikzpicture}[scale=0.65]
	
	\draw [dashed] (-6,0) -- (5,0);
	
	\draw [] (0,0.75) -- (-0.9,0.75);
	\draw  (-0.9,0.75) arc (-90:-180:1.5 and 1);
	
	\draw [] (-2.4,1.75) -- (-2.4,2.25);
	\draw [dashed] (-2.4,2.25) -- (-2.4,2.75);
	\draw [] (-2.4,2.75) -- (-2.4,3.25);
	
	\draw (-2.4,3.25) arc (90:180:3.25) ;
	
	
	\draw [] (0,0.75) -- (0,1.5);
	\draw [dashed] (0,1.5) -- (0,2.5);
	\draw [] (0,2.5) -- (0,3.25);
	\draw [] (0,3.25) -- (1.0,3.25);
	\draw [dashed] (1,3.25) -- (2,3.25);
	\draw [] (2,3.25) -- (4,3.25);
	\draw [] (4,3.25) -- (4,2.5);
	\draw [dashed] (4,1.5) -- (4,3.25);
	\draw [] (4,1.5) -- (4,0);
	
	\draw [thick,->] (0,0) -- (2.5,0) node[anchor=north west] {$x$-axis};
	\draw [thick] (0,-0.1) -- (0,0.1);
	\draw (0,0) node[anchor=north] {$x$=0};
	
	\draw [thick,<->] (-0.45,0) -- (-0.45,0.75);
	\draw (-0.45,0.375) node[anchor=east] {$D$/2};
	
	\draw [thick,<->] (0,3) -- (4,3);
	\draw (2,3) node[anchor=north] {$L_x$};

	\draw [thick,->] (4.25,0) -- (4.25,3.25);
	\draw (4.25,1.625) node[anchor=west] {$L_r$};
	
	\draw [thick,<->] (-2.4,0) -- (-4.69,2.29);
	\draw (-3.55,1.15) node[anchor=north east] {10$D$};
	
	\draw (-3.7,3.25) node[anchor=south] {$p_{high}$-chamber};
	\draw (2,3.25) node[anchor=south] {$p_{low}$-chamber};
	
	\draw [dashed] (-2.1,0) -- (-2.1,1.1);
	\draw [thick,->] (-1.5,1.6) -- (-2.1,1.2);
	\draw (-1.5,1.6) node[anchor=south] {Init.};
	
	\end{tikzpicture}%

	\caption{Schematic of the numerical domain used both in the validation case and in the test cases. The low and high pressure chamber are initially separated.}
	\label{fig:schematicDomain}
\end{figure}
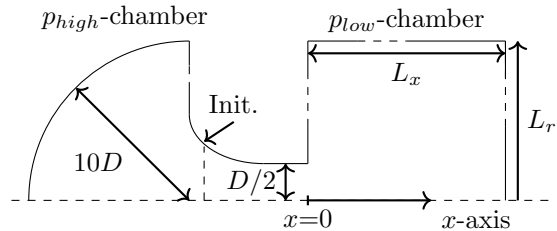

The numerical domain used for the validation case is a two degree rotationally symmetric section of a single hole injector with a diameter of $D=\SI{0.236}{mm}$ connecting the high pressure side (here: n-hexane) with the low pressure side (here: nitrogen), see Fig.~\ref{fig:schematicDomain}. The low pressure side was modeled using a constant volume chamber with $L_x/D=150$ and $L_r/D=100$. A structured mesh is used and the coordinate system is located at the nozzle exit aligned with the flow direction, starting at $x=\SI{0}{}$. Initially, both sides (low and high pressure) are separated by a membrane located at $x = -1.6D$ which bursts at simulation start resulting in the development of an expanded/underexpanded jet (depending on the prescribed pressure ratio $\Pi=p_{t,fuel}/p_{ch}$) into the positive $x$-direction. As this validation case is almost isobaric~\cite{traxinger2017a}, a low Mach number flow and a weakly expanded jet can be expected. Hence, the focus will be on the overall flow structure and the two-phase effects in the far field and a resolution of $\Delta r/D = 20$ in the radial nozzle direction was therefore chosen. The geometric specifications as well as the radial mesh resolution are summarized in Tab.~\ref{tab:schematicGeometry}. As in Traxinger \textit{et al.}~\cite{traxinger2017a}, we are using the PR-EoS for the simulation of these three test cases. This is in good agreement with our findings in section~\ref{sec:Thermodynamics} as the prediction of the PR-EoS are superior for high number alkanes like n-hexane and n-heptane compared to the SRK-EoS which gives better predictions for short alkanes, see Fig.~\ref{fig:comparisonAlkanes}. Traxinger \textit{et al.}~\cite{traxinger2017a} furthermore show that the prediction of the PR-EoS is also excellent for the binary VLE of n-hexane and nitrogen at the pressure considered for these test cases and that not fitting of the binary interaction parameter in Eq.~\eqref{eq:pseudoCritical} is necessary.

\begin{figure}[h]
	\vspace{18pt}
	\centering
	\tikzsetnextfilename{validationHexane}%
	\input{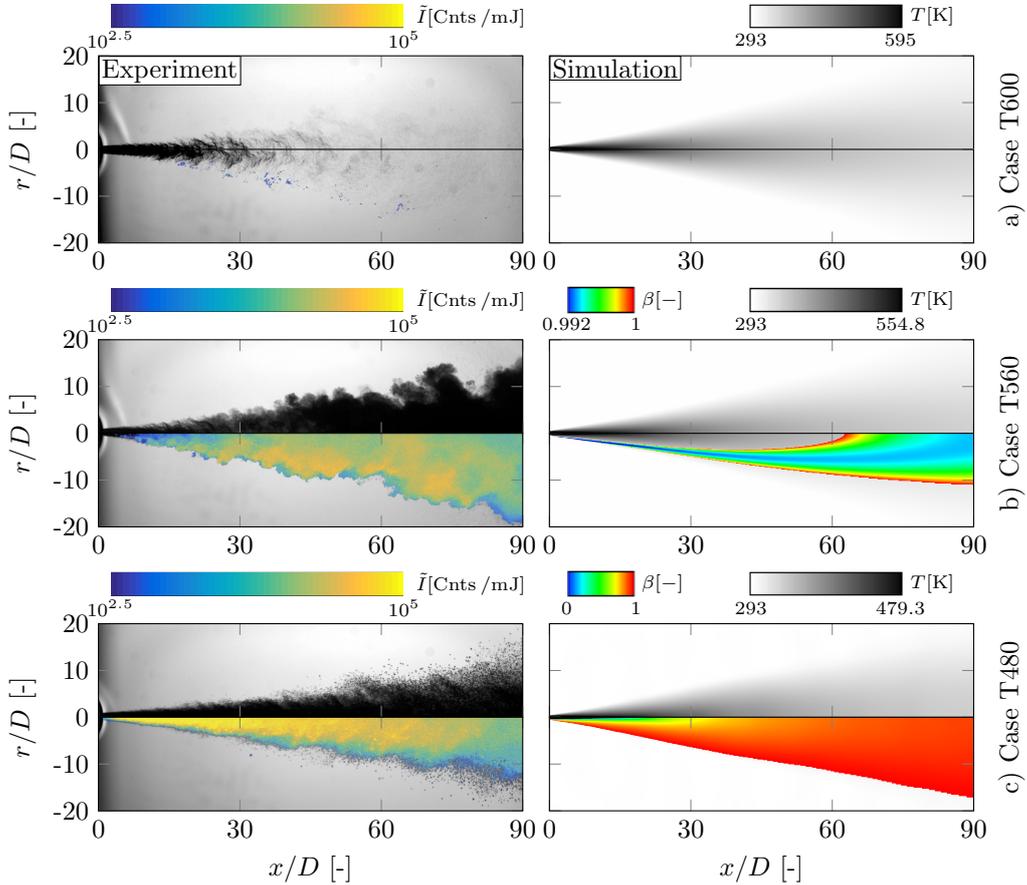}%

	\vspace{6pt}
	\caption{Comparison of experimental (left, figures taken from Traxinger \textit{et al.}~\cite{traxinger2017a}) and numerical results (right, present study) for the different test cases T600, T560 and T480.}
	\label{fig:validationHexane}
	\vspace{12pt}
\end{figure}

Figure~\ref{fig:validationHexane} shows the results of the experimental studies~\cite{traxinger2017a} on the left compared to snapshots of the numerical simulations performed in this work for the three cases T600, T560 and T480 on the right. Each frame in the left column represents a shadowgraphy image superimposed by the resulting elastic light scattering image (bottom half). In the right column the frame represents the temperature contour plot superimposed by the vapor mass fraction $\beta$ in the bottom half. For all three test cases the agreement of the numerical results with the experimental findings is remarkably good as it was also observed by Traxinger \textit{et al.}~\cite{traxinger2017a} for the conducted LESs. The experimental results for the temperature $T_{t,fuel}=\SI{600}{\kelvin}$ show no significant scattering signal and therefore Traxinger \textit{et al.}~\cite{traxinger2017a} deduced a single-phase state for this injection conditions. The same can be concluded from the numerical simulation on the right, where no region with phase separation is visible (indicated by a vapor mass fraction $\beta$ between zero and one) which in turn leads to the conclusion that the TPD-analysis was always positive for every single-phase state tested. Decreasing the total temperature to \SI{560}{\kelvin} leads to a very dark jet in the shadowgraphy image indicating the formation of dense droplet clouds. The increase in scattering intensity over some orders of magnitude confirms this statement as this measurement technique is very sensitive to particle sizes respectively (small) droplets. On the right the URANS simulation shows also a two-phase region with a vapor mass fraction of \SIrange{0.992}{1}{}. As also reported by Traxinger \textit{et al.}~\cite{traxinger2017a}, the area of phase separation in the numerical results coincides very well with regions of high scattering signal in the experiments. The third test case at $T_{t,fuel}=\SI{480}{\kelvin}$ shows a spray-like characteristic and therefore strong two-phase effects in the shadowgram as well as in the scattering. The scattering for this test case is comprising several orders of magnitude and occupies almost the entire jet domain. In addition, the shadowgram shows again a very dark jet but this time with distinct droplets visible in the outer jet region underlining the spray-like character. The numerical simulation confirms this finding. The mixture-induced phase separation is almost covering the complete jet and $\beta$ is ranging from $0$ in the inner jet region to $1$ in the outer jet areas.

\begin{table}[h]
	\centering
	\caption{Geometry specifications for the validation case and the numerical study.}
	\vspace{6pt}
	\begin{tabular}{c|cccc}
		\toprule
		\multirow{ 2}{*}{Case}  & $D$ & $L_x/D$ & $L_r/D$ & $\Delta r/D$ \\
		& [mm] 	& [-] 	& [-] 	  & [-] \\
		\midrule
		Validation 				& 0.236 & 150 	& 100	  & 20 \\
		High-pressure injection & 1.5   & 80 	& 70	  & 40 \\
		\bottomrule 
	\end{tabular}
	\label{tab:schematicGeometry}
\end{table}

All the findings for the three different test cases are in very good agreement with the LESs conducted by Traxinger \textit{et al.}~\cite{traxinger2017a}. We therefore conclude that our numerical framework consisting basically of the hybrid, pressure-based solver and the VLE-model is able to capture the main two-phase effects in weakly expanded and isobaric jets. Hence, we can use this framework to simulate highly underexpanded jets with probable phase separation.

\section{Engine-relevant Test Cases}
\label{sec:testCases}

\subsection{Test Case Description and Meshing}

The numerical investigation of the injection process of gaseous fuels at engine-relevant conditions are the focus of this study. As detailed experiments and/or simulations are lacking in literature, own setups and operating conditions were defined as follows. Based on the validation case with a chamber pressure of \SI{50}{bar} and the recent investigations of Banholzer \textit{et~al.}~\cite{banholzer2017}, who simulated high-pressure injections of hydrogen ($p_{t,fuel}=\SI{500}{bar}$) into air at rest ($p_{ch}=\SI{100}{bar}$), the test cases were chosen such that application-relevant conditions and penetration depths are met. As the purity of hydrogen is not given for industrial applications, nitrogen with a volume fraction of \SI{0.1}{\percent} is mixed with \SI{99.9}{\percent} hydrogen and defined as compressed hydrogen gas (CHG). For internal combustion engines (ICE) fired by gaseous fuels not only hydrogen is used but also compressed natural gas (CNG) with its main component methane. A literature survey showed that most of the CNG available in industry is a mixture of the three main components methane (CH$_4$), ethane (C$_2$H$_6$) and nitrogen (N$_2$)~\cite{min2002effects,kim2009effect,demirbas2010a}. Concluding the CNG-compositions available in publications a generic composition was defined, see Tab.~\ref{tab:CNGmixture}. Furthermore, the pressure levels were adapted to engine-relevant conditions. Typical chamber pressures range from \SIrange{5}{80}{\bar}, depending on the ICE-type and the start of injection (SOI). As the validation case was carried out with a chamber pressure of \SI{50}{bar} the same pressure value was chosen for the test cases. Similar considerations lead to a maximum total fuel pressure of \SI{600}{\bar}~\cite{bhaskar2013hydrogen,tuner2016combustion,tuner2016review,hu2009experimental,sierra2017working}. For a comparison of different pressure ratios additional simulations were performed with a total fuel pressure of \SI{300}{\bar}, resulting in pressure ratios $\Pi= p_{t,fuel}/p_{ch}$ of \SI{12}{} and \SI{6}{}. A total of four numerical simulations were carried out, see Tab.~\ref{tab:boundaryConditions}.

\begin{table}[!htb]
	\vspace{6pt}
	\centering
	\caption{Volume fractions [\%] of CHG- and CNG-compositions.}
	\vspace{6pt}
	\begin{tabular}{c|cccc}
		\toprule
		Fuel & Methane (CH$_4$) & Hydrogen (H$_2$) & Ethane (C$_2$H$_6$) & Nitrogen (N$_2$) \\
		\midrule
		CNG & 95.8 & 0.0 & 2.6 & 1.6 \\
		CHG & 0.0  & 99.9 & 0.0 & 0.1 \\
		\bottomrule 
	\end{tabular}
	\label{tab:CNGmixture}
	\vspace{12pt}
\end{table}

The schematic of the numerical domain used for the test cases is shown in Fig.~\ref{fig:schematicDomain} and is similar to the one for the previous validation case, except for the dimensions. For the diameter $D$  a value of \SI{1.5}{\milli\meter} was chosen, the dimensions in the axial and radial directions are $L_x/D=80$ and $L_r/D=70$. The computational setup of Hamzehloo \textit{et al.}~\cite{hamzehloo2014large,hamzehloo2016gas,hamzehloo2016numerical} and a grid independence study performed by Banholzer \textit{et al.}~\cite{banholzer2017} lead to a resolution of $\Delta x=\Delta r=D/40=\SI{0.0375}{\milli\meter}$, see Tab.~\ref{tab:schematicGeometry}.

\begin{table}[!htb]
	\vspace{6pt}
	\centering
	\caption{Overview of the boundary conditions for the four test cases.}
	\vspace{6pt}
	\begin{tabular}{c|c|ccccccc}
		\toprule
		\multirow{ 2}{*}{Case} & \multirow{ 2}{*}{EoS} & Fluid  & Fluid  & $p_{t,fuel}$ & $p_{ch}$ & $\Pi$ & $T_{t,fuel}$ & $T_{ch}$ \\
		& &  $p_{fuel}$-chamber &  $p_{ch}$-chamber & [bar] & [bar] & [-] & [K]  &[K]\\
		\midrule 
		CNG-p600 & \multirow{ 2}{*}{SRK} & \multirow{ 2}{*}{CNG} & \multirow{ 2}{*}{Nitrogen ($\textrm{N}_2$)} & 600 & 50 & 12 & \multirow{ 2}{*}{300} & \multirow{ 2}{*}{300}\\
		CNG-p300 & &  &  & 300 & 50 & 6 &  & \\
		\midrule
		CHG-p600 & \multirow{ 2}{*}{SRK} & \multirow{ 2}{*}{CHG} & \multirow{ 2}{*}{Nitrogen ($\textrm{N}_2$)} & 600 & 50 & 12 & \multirow{ 2}{*}{300} & \multirow{ 2}{*}{300}\\
		CHG-p300 & &  &  & 300 & 50 & 6 &  & \\
		\bottomrule 
	\end{tabular}
	\label{tab:boundaryConditions}
	\vspace{12pt}
\end{table}

\subsection{Probability of phase separation}

At their rest conditions both multi-component fuels CHG and CNG are in a single-phase state. As we are expecting strongly underexpanded jets due to the large prescribed pressure ratios, two different ways are feasible how the formation of a vapor-liquid equilibrium of the appropriate mixture could take place: Firstly, by an strong expansion lowering pressure and temperature at constant overall feed composition and secondly, by nitrogen dilution of the fuel and a simultaneous expansion.\\

\begin{figure}[h]
	\vspace{12pt}
	\centering
	\tikzsetnextfilename{VLEHydrogenNitrogen}%

\begin{tikzpicture}
	\begin{axis}
	[
    xmin=0,
    xmax=1,
    xlabel = {$x_{\textrm{H}_2}$, $y_{\textrm{H}_2}$, $z_{\textrm{H}_2}$ [-]},
    ymin=0,
    ymax=160,
    ylabel = $p\text{ [bar]}$,
    y filter/.code={\pgfmathparse{#1*1e-5}\pgfmathresult},
    legend style={legend pos=north west,font=\scriptsize,legend cell align=left,draw=none,fill=none}
    ]
    	\draw [color=black,line width=0.5pt,dash dot] (axis cs:0,50) -- (axis cs:1.0,50);
    	\node at (axis cs:0.85,50) [anchor=north] {\scriptsize 50 bar};
	    \addplot [color=colora,line width=1.0pt] table [x=x, y=p]{./data/criticalLocus/Hydrogen-Nitrogen/VLE_xy_Nitrogen_Hydrogen_100K.dat};
	    \addlegendentry{$T = 100.0$ K}
	    \addplot [color=colora,line width=1.0pt,forget plot] table [x=y, y=p]{./data/criticalLocus/Hydrogen-Nitrogen/VLE_xy_Nitrogen_Hydrogen_100K.dat};
 	    \addplot [color=colorb,line width=1.0pt] table [x=x, y=p]{./data/criticalLocus/Hydrogen-Nitrogen/VLE_xy_Nitrogen_Hydrogen_107.7K.dat};
 	    \addlegendentry{$T = 107.7$ K}
 	    \addplot [color=colorb,line width=1.0pt,forget plot] table [x=y, y=p]{./data/criticalLocus/Hydrogen-Nitrogen/VLE_xy_Nitrogen_Hydrogen_107.7K.dat};
		\addplot [color=colorc,line width=1.0pt] table [x=x, y=p]{./data/criticalLocus/Hydrogen-Nitrogen/VLE_xy_Nitrogen_Hydrogen_113K.dat};
		\addlegendentry{$T = 113.0$ K}
		\addplot [color=colorc,line width=1.0pt,forget plot] table [x=y, y=p]{./data/criticalLocus/Hydrogen-Nitrogen/VLE_xy_Nitrogen_Hydrogen_113K.dat};
		\addplot [color=black,line width=0.5pt,mark=*,mark size=1.0pt,only marks,fill=colora,forget plot] table [x=x, y=p]{./data/criticalLocus/Hydrogen-Nitrogen/VLE_xy_Nitrogen_Hydrogen_100K_experiment.dat};
		\addplot [color=black,line width=0.5pt,mark=*,mark size=1.0pt,only marks,fill=colora,forget plot] table [x=y, y=p]{./data/criticalLocus/Hydrogen-Nitrogen/VLE_xy_Nitrogen_Hydrogen_100K_experiment.dat}; 
		\addplot [color=black,line width=0.5pt,mark=*,mark size=1.0pt,only marks,fill=colorb,forget plot] table [x=x, y=p]{./data/criticalLocus/Hydrogen-Nitrogen/VLE_xy_Nitrogen_Hydrogen_107.7K_experiment.dat};
		\addplot [color=black,line width=0.5pt,mark=*,mark size=1.0pt,only marks,fill=colorb,forget plot] table [x=y, y=p]{./data/criticalLocus/Hydrogen-Nitrogen/VLE_xy_Nitrogen_Hydrogen_107.7K_experiment.dat};
		\addplot [color=black,line width=0.5pt,mark=*,mark size=1.0pt,only marks,fill=colorc,forget plot] table [x=x, y=p]{./data/criticalLocus/Hydrogen-Nitrogen/VLE_xy_Nitrogen_Hydrogen_113K_experiment.dat};
		\addplot [color=black,line width=0.5pt,mark=*,mark size=1.0pt,only marks,fill=colorc,forget plot] table [x=y, y=p]{./data/criticalLocus/Hydrogen-Nitrogen/VLE_xy_Nitrogen_Hydrogen_113K_experiment.dat}; 
	\end{axis}
\end{tikzpicture}%

	\vspace{6pt}
	\caption{Vapor-liquid equilibria of hydrogen and nitrogen at three different temperatures. The solid lines are calculated using the SRK-EoS. The dots are representing experimental data \cite{streett1978a,xiao1990a,shtekkel1939a}.}
	\label{fig:VLEHydrogenNitrogen}
	\vspace{12pt}
\end{figure}
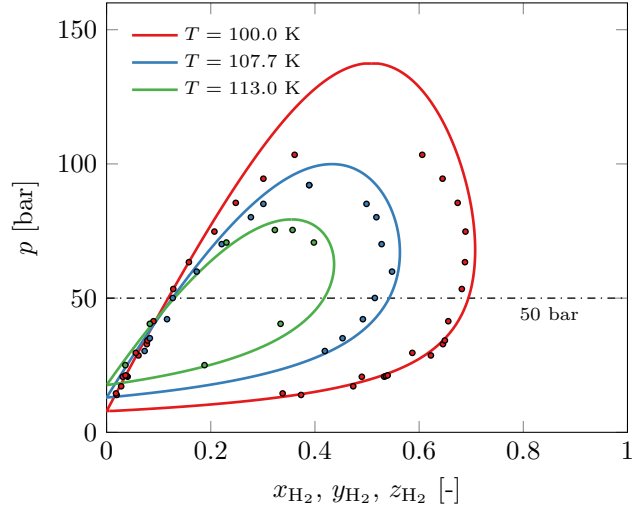

In the CHG test cases two fluids are forming the multi-component mixture, hydrogen and nitrogen, see Tab.~\ref{tab:CNGmixture}, whereby hydrogen is the low volatile and nitrogen the high volatile fluid. Hence, nitrogen is defining the upper bound of the multi-component critical locus in terms of temperature at its critical temperature \SI{126.19}{\kelvin}. Above this temperature no phase separation is possible and therefore a strong expansion and reduction of temperature is necessary to enter the multi-component VLE. In Fig.~\ref{fig:VLEHydrogenNitrogen} some selected vapor-liquid equilibria at relevant temperatures (\SI{100.0}{\kelvin}, \SI{107.7}{\kelvin} and \SI{113.0}{\kelvin}) are shown. The comparison between the experimental data and the SRK-EoS shows a reasonably good agreement, especially in the low pressure region up to the chamber pressure of \SI{50}{\bar}. Above this pressure the prediction of the SRK-EoS is getting worse with increasing pressure and decreasing temperature and the cubic EoS is overestimating the extension in terms of pressure. As CHG is almost pure hydrogen (99.9 Vol.-\%) it is obvious from Fig.~\ref{fig:VLEHydrogenNitrogen} that a phase separation at the overall feed composition of CHG can be excluded. In order to trigger a phase separation by nitrogen dilution at the chamber pressure, the mixture must contain more than 30~mole-\% nitrogen to enter the VLE at \SI{100}{\kelvin}. At the other two temperatures a minimum of approximately 40~mole-\% and 50~mole-\% $\textrm{N}_2$, respectively, has to mix into the CHG to enter the respective VLE. Hence, the probability of phase separation can be assessed low for the CHG cases.\\

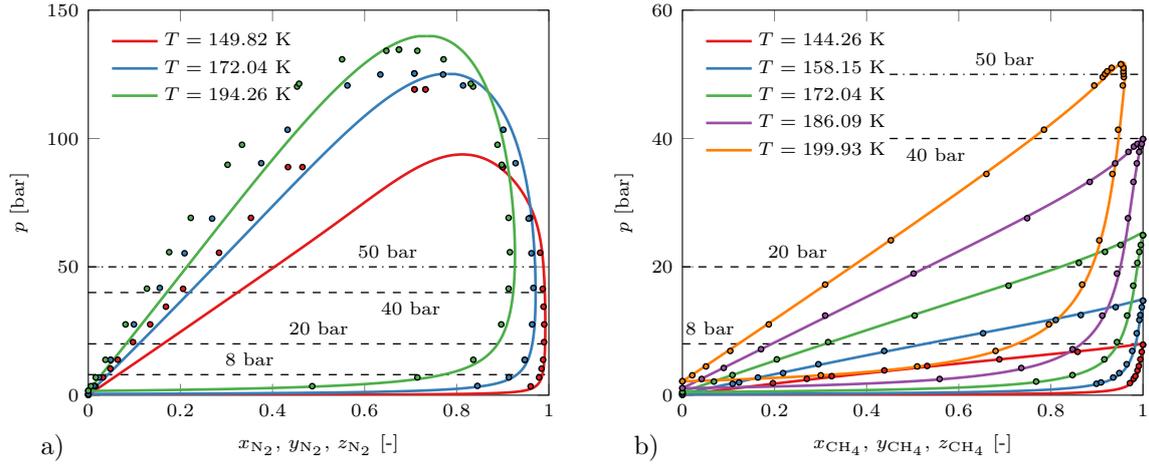
\begin{figure}[!htb]
	\vspace{12pt}
	\centering
	\tikzsetnextfilename{VLEMethaneEthane}%

\begin{tikzpicture}
	\begin{groupplot}[group style = {group name = plots,group size = 2 by 1, horizontal sep = 50pt, vertical sep = 50pt}, width = 0.48\textwidth, height = 0.42\textwidth]
	\nextgroupplot
	[
    xmin=0,
    xmax=1,
    xlabel = {$x_{\textrm{N}_2}$, $y_{\textrm{N}_2}$, $z_{\textrm{N}_2}$ [-]},
    x filter/.code={\pgfmathparse{1-#1}\pgfmathresult},
    ymin=0,
    ymax=150,
    ylabel = $p\text{ [bar]}$,
    y filter/.code={\pgfmathparse{#1*1e-5}\pgfmathresult},
    legend style={legend pos=north west,font=\scriptsize,legend cell align=left,draw=none,fill=none},
    yticklabel style = {font=\scriptsize},
    xticklabel style = {font=\scriptsize},
    ylabel style = {font=\scriptsize},
    xlabel style = {font=\scriptsize}
    ]
    	\draw [color=black,line width=0.5pt,dash dot] (axis cs:0,50) -- (axis cs:1.0,50);
    	\node at (axis cs:0.65,50) [anchor=south] {\scriptsize 50 bar};
    	\draw [color=black,line width=0.5pt,dashed] (axis cs:0,40) -- (axis cs:1,40);
    	\node at (axis cs:0.7,40) [anchor=north] {\scriptsize 40 bar};
    	\draw [color=black,line width=0.5pt,dashed] (axis cs:0,20) -- (axis cs:1,20);
 	    \node at (axis cs:0.5,20) [anchor=south] {\scriptsize 20 bar};
 	    \draw [color=black,line width=0.5pt,dashed] (axis cs:0,8) -- (axis cs:1,8);
 	    \node at (axis cs:0.35,8) [anchor=south] {\scriptsize 8 bar};
	    \addplot [color=colora,line width=1.0pt] table [x=x, y=p]{./data/criticalLocus/Ethane-Nitrogen/VLE_xy_Nitrogen_Ethane_149.82K.dat};
	    \addlegendentry{$T = 149.82$ K}
	    \addplot [color=colora,line width=1.0pt,forget plot] table [x=y, y=p]{./data/criticalLocus/Ethane-Nitrogen/VLE_xy_Nitrogen_Ethane_149.82K.dat};
 	    \addplot [color=colorb,line width=1.0pt] table [x=x, y=p]{./data/criticalLocus/Ethane-Nitrogen/VLE_xy_Nitrogen_Ethane_172.04K.dat};
 	    \addlegendentry{$T = 172.04$ K}
 	    \addplot [color=colorb,line width=1.0pt,forget plot] table [x=y, y=p]{./data/criticalLocus/Ethane-Nitrogen/VLE_xy_Nitrogen_Ethane_172.04K.dat};
  	    \addplot [color=colorc,line width=1.0pt] table [x=x, y=p]{./data/criticalLocus/Ethane-Nitrogen/VLE_xy_Nitrogen_Ethane_194.26K.dat};
  	    \addlegendentry{$T = 194.26$ K}
  	    \addplot [color=colorc,line width=1.0pt,forget plot] table [x=y, y=p]{./data/criticalLocus/Ethane-Nitrogen/VLE_xy_Nitrogen_Ethane_194.26K.dat};
 	    \addplot [color=black,line width=0.5pt,mark=*,mark size=1.0pt,only marks,fill=colora,forget plot] table [x=x, y=p]{./data/criticalLocus/Ethane-Nitrogen/VLE_xy_Nitrogen_Ethane_149.82K_experiment.dat};
 	    \addplot [color=black,line width=0.5pt,mark=*,mark size=1.0pt,only marks,fill=colora,forget plot] table [x=y, y=p]{./data/criticalLocus/Ethane-Nitrogen/VLE_xy_Nitrogen_Ethane_149.82K_experiment.dat};
   	    \addplot [color=black,line width=0.5pt,mark=*,mark size=1.0pt,only marks,fill=colorb,forget plot] table [x=x, y=p]{./data/criticalLocus/Ethane-Nitrogen/VLE_xy_Nitrogen_Ethane_172.04K_experiment.dat};
   	    \addplot [color=black,line width=0.5pt,mark=*,mark size=1.0pt,only marks,fill=colorb,forget plot] table [x=y, y=p]{./data/criticalLocus/Ethane-Nitrogen/VLE_xy_Nitrogen_Ethane_172.04K_experiment.dat};
   	    \addplot [color=black,line width=0.5pt,mark=*,mark size=1.0pt,only marks,fill=colorc,forget plot] table [x=x, y=p]{./data/criticalLocus/Ethane-Nitrogen/VLE_xy_Nitrogen_Ethane_194.26K_experiment.dat};
   	    \addplot [color=black,line width=0.5pt,mark=*,mark size=1.0pt,only marks,fill=colorc,forget plot] table [x=y, y=p]{./data/criticalLocus/Ethane-Nitrogen/VLE_xy_Nitrogen_Ethane_194.26K_experiment.dat};
	\nextgroupplot
	[
	xmin=0,
	xmax=1,
	xlabel = {$x_{\textrm{C}\textrm{H}_4}$, $y_{\textrm{C}\textrm{H}_4}$, $z_{\textrm{C}\textrm{H}_4}$ [-]},
	ymin=0,
	ymax=60,
	ylabel = $p\text{ [bar]}$,
	y filter/.code={\pgfmathparse{#1*1e-5}\pgfmathresult},
	legend style={legend pos=north west,font=\scriptsize,legend cell align=left,draw=none,fill=none},
	yticklabel style = {font=\scriptsize},
	xticklabel style = {font=\scriptsize},
	ylabel style = {font=\scriptsize},
	xlabel style = {font=\scriptsize}
	]
		\draw [color=black,line width=0.5pt,dash dot] (axis cs:0.45,50) -- (axis cs:1.0,50);
		\node at (axis cs:0.70,50) [anchor=south] {\scriptsize 50 bar};
    	\draw [color=black,line width=0.5pt,dashed] (axis cs:0.45,40) -- (axis cs:1,40);
    	\node at (axis cs:0.55,40) [anchor=north] {\scriptsize 40 bar};
    	\draw [color=black,line width=0.5pt,dashed] (axis cs:0,20) -- (axis cs:1,20);
		\node at (axis cs:0.25,20) [anchor=south] {\scriptsize 20 bar};
		\draw [color=black,line width=0.5pt,dashed] (axis cs:0,8) -- (axis cs:1,8);
		\node at (axis cs:0.06,8) [anchor=south] {\scriptsize 8 bar};
		\addplot [color=colora,line width=1.0pt] table [x=x, y=p]{./data/criticalLocus/Methane-Ethane/VLE_xy_Ethane_Methane_144.26K.dat};
		\addlegendentry{$T = 144.26$ K}
		\addplot [color=colora,line width=1.0pt,forget plot] table [x=y, y=p]{./data/criticalLocus/Methane-Ethane/VLE_xy_Ethane_Methane_144.26K.dat};
		\addplot [color=colorb,line width=1.0pt] table [x=x, y=p]{./data/criticalLocus/Methane-Ethane/VLE_xy_Ethane_Methane_158.15K.dat};
		\addlegendentry{$T = 158.15$ K}
		\addplot [color=colorb,line width=1.0pt,forget plot] table [x=y, y=p]{./data/criticalLocus/Methane-Ethane/VLE_xy_Ethane_Methane_158.15K.dat};
		\addplot [color=colorc,line width=1.0pt] table [x=x, y=p]{./data/criticalLocus/Methane-Ethane/VLE_xy_Ethane_Methane_172.04K.dat};
		\addlegendentry{$T = 172.04$ K}
		\addplot [color=colorc,line width=1.0pt,forget plot] table [x=y, y=p]{./data/criticalLocus/Methane-Ethane/VLE_xy_Ethane_Methane_172.04K.dat};
		\addplot [color=colord,line width=1.0pt] table [x=x, y=p]{./data/criticalLocus/Methane-Ethane/VLE_xy_Ethane_Methane_186.09K.dat};
		\addlegendentry{$T = 186.09$ K}
		\addplot [color=colord,line width=1.0pt,forget plot] table [x=y, y=p]{./data/criticalLocus/Methane-Ethane/VLE_xy_Ethane_Methane_186.09K.dat};
		\addplot [color=colore,line width=1.0pt] table [x=x, y=p]{./data/criticalLocus/Methane-Ethane/VLE_xy_Ethane_Methane_199.93K.dat};
		\addlegendentry{$T = 199.93$ K}
		\addplot [color=colore,line width=1.0pt,forget plot] table [x=y, y=p]{./data/criticalLocus/Methane-Ethane/VLE_xy_Ethane_Methane_199.93K.dat};
	    \addplot [color=black,line width=0.5pt,mark=*,mark size=1.0pt,only marks,fill=colora,forget plot] table [x=x, y=p]{./data/criticalLocus/Methane-Ethane/VLE_xy_Ethane_Methane_144.26K_experiment.dat};
	    \addplot [color=black,line width=0.5pt,mark=*,mark size=1.0pt,only marks,fill=colora,forget plot] table [x=y, y=p]{./data/criticalLocus/Methane-Ethane/VLE_xy_Ethane_Methane_144.26K_experiment.dat};
 	    \addplot [color=black,line width=0.5pt,mark=*,mark size=1.0pt,only marks,fill=colorb,forget plot] table [x=x, y=p]{./data/criticalLocus/Methane-Ethane/VLE_xy_Ethane_Methane_158.15K_experiment.dat};
 	    \addplot [color=black,line width=0.5pt,mark=*,mark size=1.0pt,only marks,fill=colorb,forget plot] table [x=y, y=p]{./data/criticalLocus/Methane-Ethane/VLE_xy_Ethane_Methane_158.15K_experiment.dat};
  	    \addplot [color=black,line width=0.5pt,mark=*,mark size=1.0pt,only marks,fill=colorc,forget plot] table [x=x, y=p]{./data/criticalLocus/Methane-Ethane/VLE_xy_Ethane_Methane_172.04K_experiment.dat};
  	    \addplot [color=black,line width=0.5pt,mark=*,mark size=1.0pt,only marks,fill=colorc,forget plot] table [x=y, y=p]{./data/criticalLocus/Methane-Ethane/VLE_xy_Ethane_Methane_172.04K_experiment.dat};
   	    \addplot [color=black,line width=0.5pt,mark=*,mark size=1.0pt,only marks,fill=colord,forget plot] table [x=x, y=p]{./data/criticalLocus/Methane-Ethane/VLE_xy_Ethane_Methane_186.09K_experiment.dat};
   	    \addplot [color=black,line width=0.5pt,mark=*,mark size=1.0pt,only marks,fill=colord,forget plot] table [x=y, y=p]{./data/criticalLocus/Methane-Ethane/VLE_xy_Ethane_Methane_186.09K_experiment.dat};
   	    \addplot [color=black,line width=0.5pt,mark=*,mark size=1.0pt,only marks,fill=colore,forget plot] table [x=x, y=p]{./data/criticalLocus/Methane-Ethane/VLE_xy_Ethane_Methane_199.93K_experiment.dat};
   	    \addplot [color=black,line width=0.5pt,mark=*,mark size=1.0pt,only marks,fill=colore,forget plot] table [x=y, y=p]{./data/criticalLocus/Methane-Ethane/VLE_xy_Ethane_Methane_199.93K_experiment.dat};
    \end{groupplot}
   	\node[below = 0.4cm of plots c1r1.south,xshift=-3.5cm] {a)};
   	\node[below = 0.4cm of plots c2r1.south,xshift=-3.5cm] {b)};
\end{tikzpicture}%

	\vspace{6pt}
	\caption{VLEs for binary mixtures: a) nitrogen + ethane and b) methane + ethane. The solid lines are calculated using the SRK-EoS. The dots represent experimental data \cite{stryjek1974a,wichterle1972a}.}
	\label{fig:VLEMethaneEthane}
	\vspace{12pt}
\end{figure}

For the CNG the VLE and the probability of phase separation is significantly different. Firstly, CNG is defined as a ternary mixture consisting of methane, ethane and nitrogen, see Tab.~\ref{tab:CNGmixture}, and therefore the mixture space is increased by one dimension compared to binary mixtures. Secondly, nitrogen is now the low volatile component because methane and ethane are having a significantly higher critical temperature, see Tab.~\ref{tab:criticalProperties}. Therefore, the critical locus has in the case of CNG an upper bound at 305.32~K which is the critical temperature of ethane. This temperature in turn is almost three times larger than the critical temperature of nitrogen which is the high volatile component in the CHG mixture.
\begin{figure}[p]
	\centering
	\tikzsetnextfilename{ternaryMixtures}%
	\input{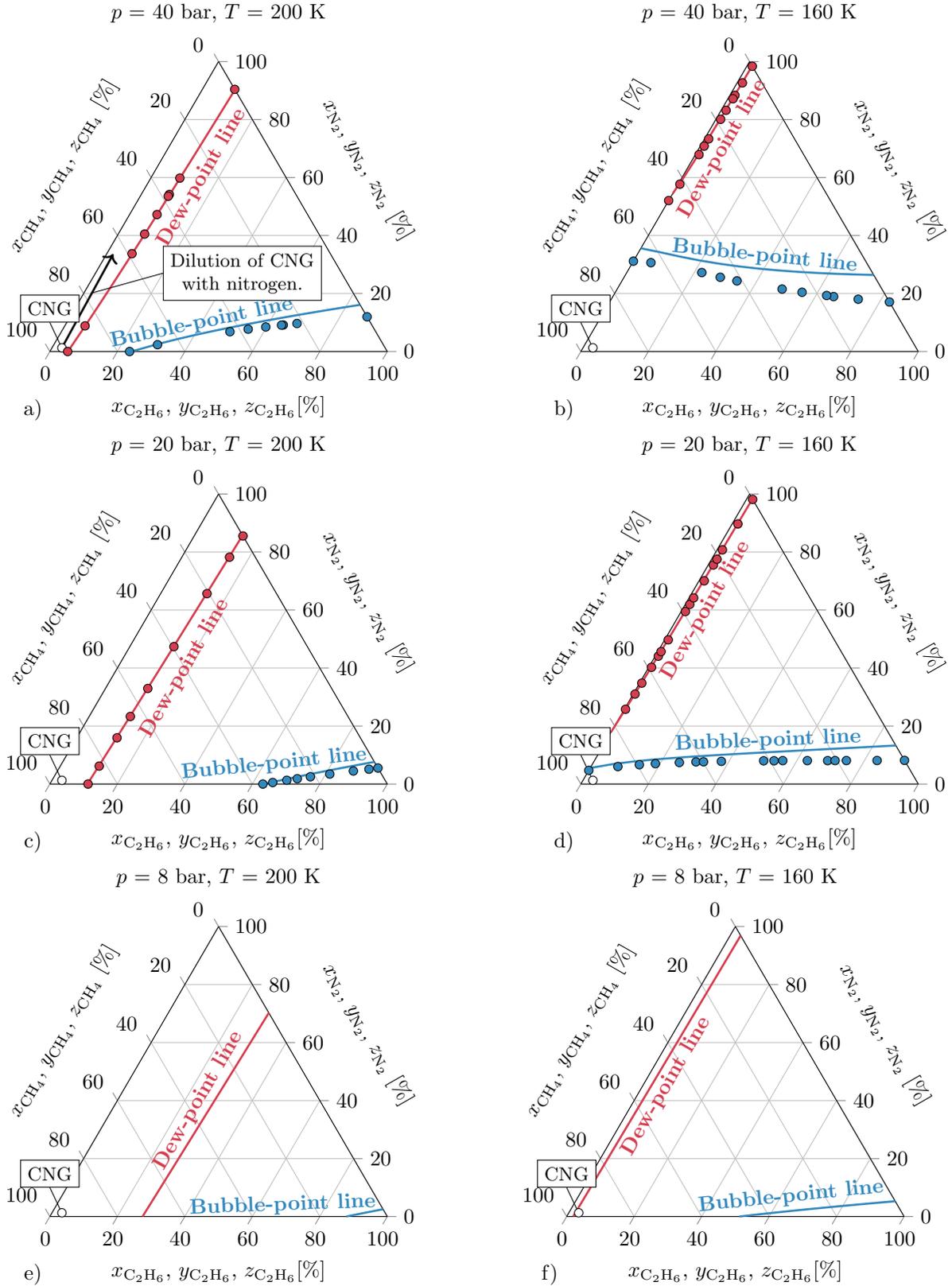}%

	\vspace{-6pt}
	\caption{VLEs for the ternary mixture consisting of methane, ethane and nitrogen at two different temperatures (160 K and 200 K) and three different pressures (40 bar, 20 bar and 8 bar). The CNG mixture as defined in Tab.~\ref{tab:CNGmixture} is marked with a white dot. The solid lines are calculated using the SRK-EoS. The dots are representing experimental data \cite{trappehl1987a}.}
	\label{fig:ternaryMixtures}
\end{figure}
The higher maximum critical temperature makes phase separation during the CNG injection significantly more probable than in the CHG-case because the injection temperature and the critical temperature of ethane are almost equal (Recall: For the CHG-case an expansion below 126.19~K is at least necessary to enter the binary VLE.). Figure~\ref{fig:ternaryMixtures} shows VLEs of the ternary mixture at three different pressures (40~bar, 20~bar and 8~bar) and two different temperatures (200~K and 160~K). The dew-point lines are shown in red and the bubble-point lines in blue while the experiments are represented by dots and the solid lines are calculated using the SRK-EoS. The feed composition of CNG is marked by the white dot in each plot. In order to give a more detailed glance onto these ternary VLE-diagrams, we also plotted the related binary mixtures into $pxy$-diagrams, see Fig.~\ref{fig:criticalLocus}~c for nitrogen + methane, Fig.~\ref{fig:VLEMethaneEthane}~a for nitrogen + ethane and Fig.~\ref{fig:VLEMethaneEthane}~b for methane + ethane. In the appropriate diagrams the three investigated pressure levels (8~bar, 20~bar and 40~bar) are marked with dashed lines and the chamber pressure of 50~bar is marked with a dashed dotted line. Especially in Fig.~\ref{fig:VLEMethaneEthane}~a, the deviations between the experimental data and the cubic EoS at the bubble-point line become very obvious, where cubic EoS are always facing considerable inaccuracies. With decreasing temperature the prediction error is rising steadily. Similar findings can be made for the binary mixture nitrogen + methane, but compared to the nitrogen-ethane mixture the overall prediction accuracy is acceptable over the investigated pressure and temperature range, compare also the very good prediction of the critical locus in Fig.~\ref{fig:criticalLocus}~d. An excellent prediction is evident for the binary mixture methane + ethane over the relevant pressure and temperature range. These findings for the different binary mixtures transfer directly to the VLEs of the ternary mixture CNG. Apart from the \SI{40}{\bar} and \SI{160}{\kelvin} diagram, see Fig.~\ref{fig:ternaryMixtures}~b, the comparison to experimental data is very good, see Figs.~\ref{fig:ternaryMixtures}~a, c and d. Especially the dew-point line is predicted excellently by the SRK-EoS and the already mentioned problems in terms of prediction accuracy are obvious at the bubble-point line. This in turn is not a big issue in this investigation as most of our two-phase states will occupy the left corner of the ternary VLE-diagram, where the prediction of the SRK-EoS is very good. This can be seen in detail in Fig.~\ref{fig:ternaryMixtures}~a where the feed composition of the CNG and the dilution process of CNG with nitrogen are shown. As in our case the VLE will be entered across the dew-point line due to the strong expansion process, we can expect an excellent prediction concerning the onset of phase separation.

\subsection{Choked nozzle flow}

Using the one-dimensional flow analysis presented in Banholzer \textit{et~al.}~\cite{banholzer2017} the choked flow properties of the two fuels for the different operating conditions can be calculated, see Tab.~\ref{tab:critNozzle}. The larger critical pressure $p^\ast$ of CHG leads to an earlier choking of the nozzle compared to the CNG cases. As we are using a constant chamber pressure of 50 bar for all four test cases the flow in the nozzle is always choked. For this type of flow the maximum mass flow rate through the orifice can be evaluated as
\begin{equation}
\dot{m}_n^\ast = \rho^\ast a_s^\ast A
\end{equation}
with $A$ being the cross-sectional area of the nozzle. Reduction of the cross section of the vena contracta due to detachments caused by the shape of the nozzle and friction effects inside the nozzle can decrease the mass flow by up to \SI{20}{\percent} \cite{tang1978experimental}. The reduction of the achievable mass flow is known as discharge coefficient and is defined as $C_d = \dot{m} / \dot{m}_n^\ast$. For the lower fuel pressure of \SI{300}{\bar} the maximum mass flow of CNG is $3.6$-times higher than for CHG, mainly due to the approximately ten times higher density $\rho^\ast$ of CNG compared to CHG. Doubling the pressure to \SI{600}{\bar} leads to an increase of the critical mass flow rate of nearly a factor of 2 for both fuels.

\begin{table}[!htb]
	\vspace{12pt}
	\centering
	\caption{Choked nozzle flow properties for CHG and CNG ($T_{t,fuel}=\SI{300}{\kelvin}$, SRK-EoS).}
	\vspace{6pt}
	\begin{tabular}{c|c|ccccc}
		\toprule
		\multirow{ 2}{*}{Fuel}  & $p_{t,fuel}$ & $p^\ast$ & $\rho^\ast$ & $a_s^\ast$ & $\dot{m}_n^\ast$ & $\dot{M}_n^\ast$ \\
								& [\si{bar}] & [\si{bar}] 	& [\si{\kilogram\per\meter\cubed}] 	& [\si{\meter\per\second}] 	  & [\si{\gram\per\second}] 	  & [\si{\kilogram\meter\per\second\squared}] \\
		\midrule
		\multirow{ 2}{*}{CNG}	& 600 & 191.37 & 220.55 & 556.42 & 216.86 & 120.67 \\
		    					& 300 & 124.66 & 148.87 & 439.36 & 115.59 & 50.78 \\
		\midrule
		\multirow{ 2}{*}{CHG}   & 600 & 279.61 & 23.87 & 1469.18 & 62.02 & 91.11 \\
			 					& 300 & 148.09 & 13.65 & 1330.11 & 32.09 & 42.68 \\
		\bottomrule 
	\end{tabular}
	\label{tab:critNozzle}
	\vspace{18pt}
\end{table}

The main influencing parameter of the penetration depth of the jet is the momentum flux being present at the nozzle exit \cite{abraham1996entrapment,hill1999transient,gerold2013new}. The momentum flux in a choked flow is defined as:
\begin{equation}
\dot{M}_n^\ast = \rho^\ast {a_s^\ast}^2 A \; .
\end{equation}
For the lower fuel pressure of \SI{300}{\bar} the value evaluated for the CNG-case is higher than for the CHG-case. The difference of the two fluids is not as big as in case of the mass flux because the momentum flux is proportional to the squared speed of sound. For the fuel pressure of \SI{600}{\bar} the momentum flux increases by approximately a factor of 2. Comparing the values for both fuels and operating points it is presumed that the CNG-jet with \SI{600}{\bar} (CNG-p600) penetrates the fastest, followed by CHG-p600, CNG-p300 and CHG-p300.

\section{Results}
\label{sec:results}

\subsection{Detailed Discussion: CNG-p600}

\begin{figure}[h]
	\vspace{12pt}
	\centering
	\tikzsetnextfilename{CNGp600_temporalEvolution}%
	\input{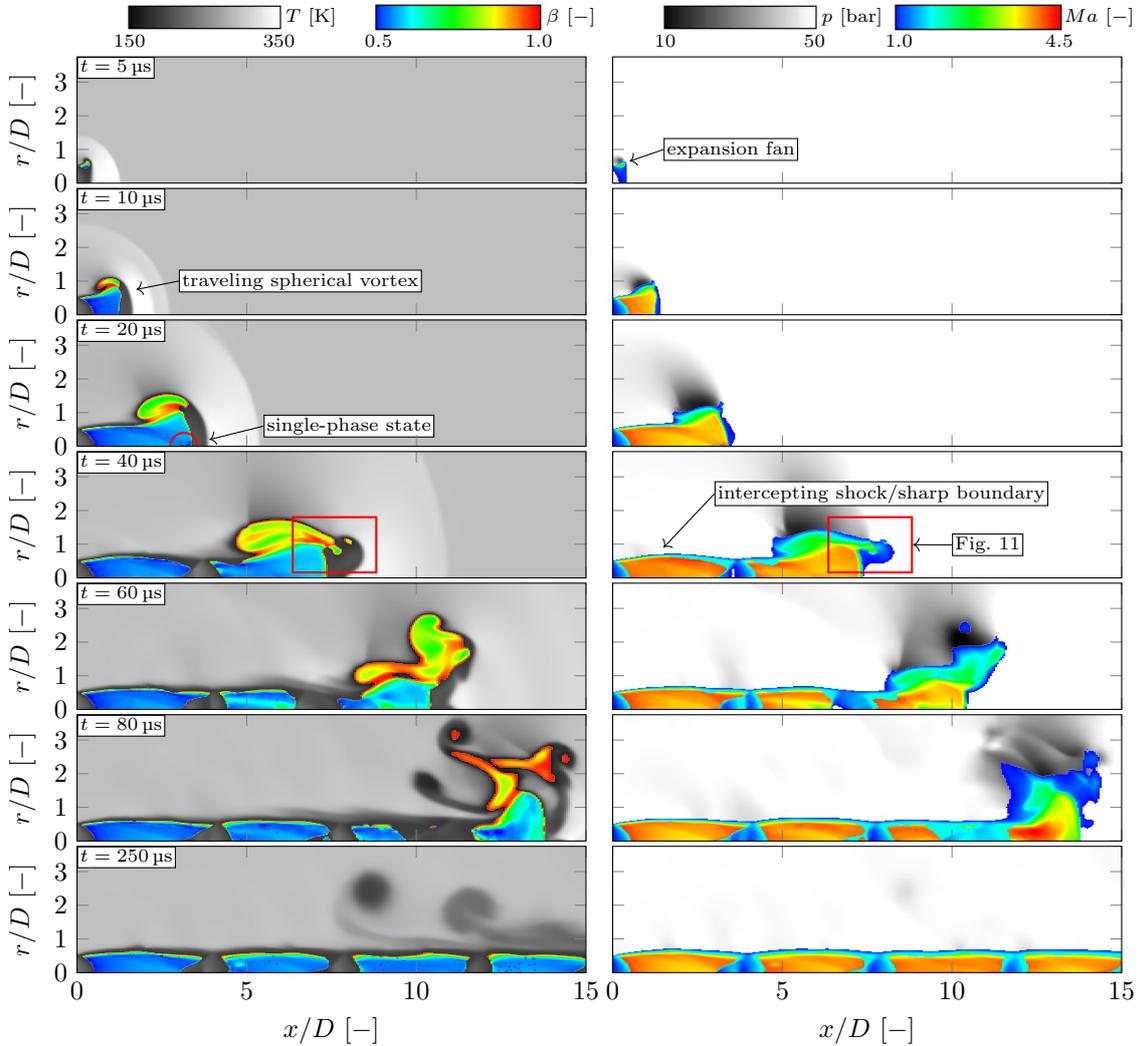}%

	\vspace{6pt}
	\caption{Snapshots of the temporal evolution of the underexpanded jet for the test case CNG-p600 ($p_{t,fuel}=\SI{600}{\bar}$). In the left column the temperature and the vapor mass fraction $\beta$ are superimposed to visualize regions where the jet enters the VLE. In the right column the superposition of pressure and Mach number is shown to get a more detailed glance of the supersonic flow accompanying the jet.}
	\label{fig:CNGp600_temporalEvolution}
	\vspace{12pt}
\end{figure}

Fig.~\ref{fig:CNGp600_temporalEvolution} shows snapshots of the temporal evolution of the high-pressure injection of CNG with a total pressure of \SI{600}{\bar} and a total temperature of \SI{300}{\kelvin} into the pressurized chamber filled with nitrogen at rest ($p_{ch}=\SI{50}{\bar}, T_{ch}=\SI{300}{\kelvin}$). For the detailed discussion we are only showing the first \SI{20}{\percent} of the simulated chamber which is in our opinion sufficient to discuss all the main fluid-dynamic and thermodynamic effects happening during the transient part of the injection process. The left column represents the temperature distribution superimposed by the vapor mass fraction $\beta$. In the right column the pressure field is superimposed by the Mach number, in which regions with $\textrm{Ma}<1$ are excluded.\\

Looking at the overall flow structure in the different snapshots one can see the typical flow phenomena occurring in underexpanded jets. The static temperature as well as the static pressure are covering a wide range whereby the large values are caused by the compression process downstream of the tip of the jet and the low values are related to the strong expansion within the jet. For a better visualization we bounded the upper limit of the pressure to 50 bar. After the start of injection (\SI{5}{\micro\second}) the fluid enters the mixing chamber and a Prandtl-Mayer expansion fan is formed at the nozzle edge. The flow accelerates across the expansion fan and therefore the Mach number increases while the static pressure and temperature drop inside the jet. As the jet penetrates further into the chamber a traveling spherical vortex forms in front of the jet. This spherical vortex is growing steadily and breaks up at approximately \SI{60}{\micro\second}. The fluid inside this vortex gets pushed into the mixing chamber by the subsequent steady-state jet region which is apparent in the Mach number plots at $t = \SI{60}{\micro\second}$ and $t = \SI{80}{\micro\second}$. Apart from this dominant vortex structure shock phenomena can be observed during the injection process which can be seen for the first time at \SI{40}{\micro\second}. The expansion fans being present at the outer part of the jet interact with the free jet boundary leading to weak compression waves and an intercepting shock, which distinguishes the jet from the surrounding nitrogen. This intercepting shock in turn ends in a strong curved shock, the Mach disk being a very prominent feature in such underexpanded jets. In our case, this concise shock is rather weak/oblique and therefore the flow remains at supersonic conditions ($\textrm{Ma}>1$). Downstream of this shock the flow accelerates quickly again due to the large pressure ratios being present in the underexpanded jet until another Mach disk forms and decelerates the flow again. As time moves on and the jet penetrates further through the chamber, several of this shock structures called shock barrels form along the jet. The last row in Fig.~\ref{fig:CNGp600_temporalEvolution} shows the quasi-stationary jet after an injection time of \SI{250}{\micro\second} where a total of four shock barrels is visible and all flow properties are only fluctuating around their mean values.\\

\begin{figure}[t]
	\centering
	\tikzsetnextfilename{CNGp600_snaps}%
	\input{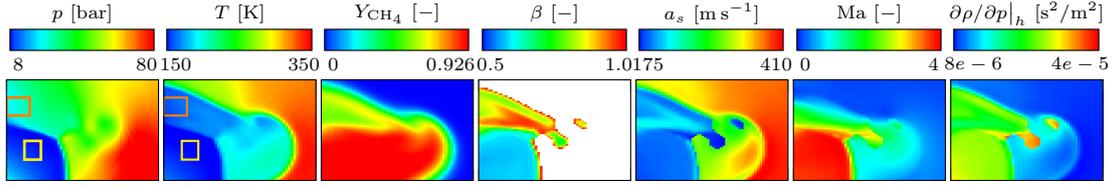}%

	\vspace{6pt}
	\caption{Detailed view of different flow properties for the CNG-p600 test case at \SI{40}{\micro\second} after injection. Snapshots are taken at the tip of the jet including both the jet center as well as the traveling spherical vortex, for more details see Fig.~\ref{fig:CNGp600_temporalEvolution} red box. The yellow and orange boxes mark the sampling points for Fig.~\ref{fig:tenaryMixturesDetail}.}
	\label{fig:CNGp600_snaps}
\end{figure}

From the left column in Fig.~\ref{fig:CNGp600_temporalEvolution} it is obvious that larger parts of the jet have entered the VLE and therefore the underexpanded jet features pronounced two-phase effects which can be identified from the vapor mass fraction $\beta$ ranging from \SI{0.5}{} to \SI{1}{}. At \SI{5}{\micro\second} a first small region with phase separation has formed very close to the nozzle trailing edge resulting from the decrease of the static pressure and temperature inside the Prandtl-Mayer expansion fan. Related to this is a discontinuous drop in the speed of sound and in turn a non-linear increase in the Mach number from approximately unity to Mach numbers ranging from 2 up to 3. The subsequent expansion leads to a further decrease of pressure and temperature and at \SI{10}{\micro\second} a large part of the jet has entered the VLE. This trend continues until the formation of a first shock barrel sets in. The start of the formation of this shock structure can be seen in terms of a small single-phase region close to the centerline at $t = \SI{20}{\micro\second}$. At $t = \SI{40}{\micro\second}$ the first shock barrel is already fully developed and can be seen at $x/D < 4$. Due to the shock and the accompanying increase in temperature and pressure the jet gets back to a single-phase state upstream of the shock barrel. As the flow is accelerated after this shock, the continuous expansion leads again to a phase instability and therefore to a phase separation. This phenomenon happens several times during the propagation of the jet into the chamber and leads to four distinct shock barrels containing one large region of two-phase flow each at $t = \SI{250}{\micro\second}$. Outside of this region with strong expansion large areas of phase separation are also visible in the traveling spherical vortex. Compared to the jet center, the vapor mass fraction is larger and more inhomogeneous, i.e. covering a wider range. This fact can be explained by taking a deeper look in the phase separation process itself and how it is triggered in these two different regions of the jet. This will be done with results from the snapshot at $t = \SI{40}{\micro\second}$, where we will focus on a small region at the jet tip marked by the red box in Fig.~\ref{fig:CNGp600_temporalEvolution} where both phase separation regions are present (jet center and traveling spherical vortex). In Fig.~\ref{fig:CNGp600_snaps} the detailed view of this region is visualized in terms of the static pressure $p$, the static temperature $T$, the methane mass fraction $Y_{CH_4}$, the vapor mass fraction $\beta$, the speed of sound $a_s$, the Mach number $\textrm{Ma}$ and the partial derivative $\partial \rho / \partial p \at[\big]{h}$.

\begin{figure}[h]
	\vspace{6pt}
	\centering
	\tikzsetnextfilename{ternaryMixtures_Detail}%
	\input{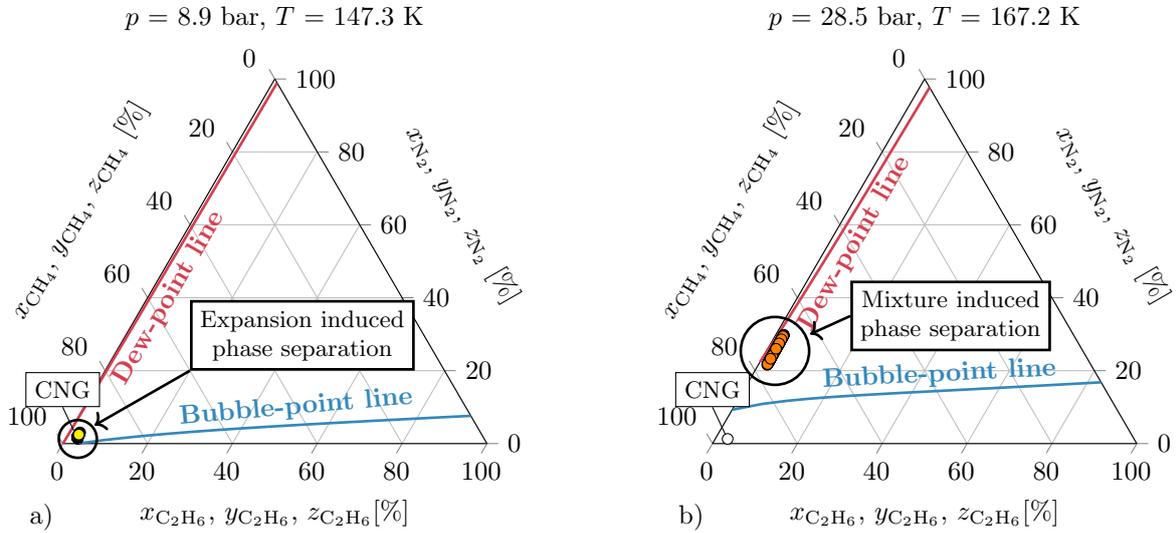}%

	\vspace{6pt}
	\caption{Ternary phase diagrams for the CNG test-case CNG-p600. The colored points mark the sampling data highlighted in Fig.~\ref{fig:CNGp600_snaps} by the appropriate colored boxes. The solid lines as well as the sampling data for the numerical simulations at \SI{40}{\micro\second} are both calculated with the SRK-EoS.}
	\label{fig:tenaryMixturesDetail}
	\vspace{6pt}
\end{figure}

Beginning on the left, the first snapshot reveals the large pressure gradient being present at the tip of the underexpanded jet. The minimum pressure of the first expansion wave is approximately \SI{8}{\bar} while the compression process in front of the jet increases the fluid-pressure up to \SI{80}{\bar} being 1.6 times larger than the prescribed chamber pressure of \SI{50}{\bar}. Due to this strong expansion and compression, the static temperature ranges approximately from \SI{150}{\kelvin} to \SI{350}{\kelvin} (second frame) showing a pattern similar to the static pressure. In the third frame the mass fraction of methane is shown which is the main component of CNG, see Tab.~\ref{tab:CNGmixture}. At \SI{40}{\micro\second} almost no nitrogen has mixed into the jet core yet which is therefore almost at the feed composition. In the region of the traveling spherical vortex the opposite is true as the convection inside of this vortex is forcing the dilution of CNG with the surrounding nitrogen. As a consequence, two different phase separation regions can be distinguished during the CNG injection process, a phase separation solely caused by expansion and a phase separation triggered by expansion and nitrogen dilution. In order to emphasize this finding we have sampled two appropriate areas of data points at the positions marked by the yellow and orange box in Fig.~\ref{fig:CNGp600_snaps} and plotted them in ternary mixture diagrams, see Fig.~\ref{fig:tenaryMixturesDetail}.
The constant temperature and pressure in these diagrams were chosen such that a large number of meaningful points can be plotted. In Fig.~\ref{fig:tenaryMixturesDetail}~a) the scattering data of the expansion induced phase separation (yellow) is plotted together with the VLE. Almost no mixing with nitrogen has taken place as already discussed and therefore all data points are grouping around the feed composition. Due to the strong expansion occurring at the jet center the VLE is almost occupying the complete mixture space and not surprisingly all points in the CFD have entered the region of single-phase instability. For the sample data taken from the traveling spherical vortex the sole expansion process would not have been sufficient to lead to a phase separation. From Fig.~\ref{fig:tenaryMixturesDetail}~b) it gets obvious that a dilution of the CNG with the surrounding nitrogen is necessary to enter the VLE at the appropriate temperature and pressure. Due to the convective mixing inside the vortex the individual mixtures contain more than 20 mole-\% nitrogen and are therefore able to enter the VLE. Going back to Fig.~\ref{fig:CNGp600_snaps} one can now reconstruct the snapshot for the vapor mass fraction $\beta$ shown in the fourth frame. In the jet center the variation in temperature, pressure and composition is low and therefore an almost constant $\beta$-value can be observed. Inside the vortex a wider range of $\beta$-values can be seen due to the steady dilution/mixing with the surrounding nitrogen and also the wider pressure and temperature range. As already discussed earlier, the speed of sound $a_s$ (fifth frame), the Mach number Ma (sixth frame) and the partial derivative $\partial \rho / \partial p \at[\big]{h}$ exhibit a discontinuous jump as the mixture enters the VLE. The speed of sound drops significantly and the Mach number defined as $\textrm{Ma}=u/a_s$ increases to values up to \SI{4}{}.

\subsection{Effects of pressure change for CNG}

\begin{figure}[t]
	\vspace{6pt}
	\centering
	\tikzsetnextfilename{CNG_comparison}%
	\input{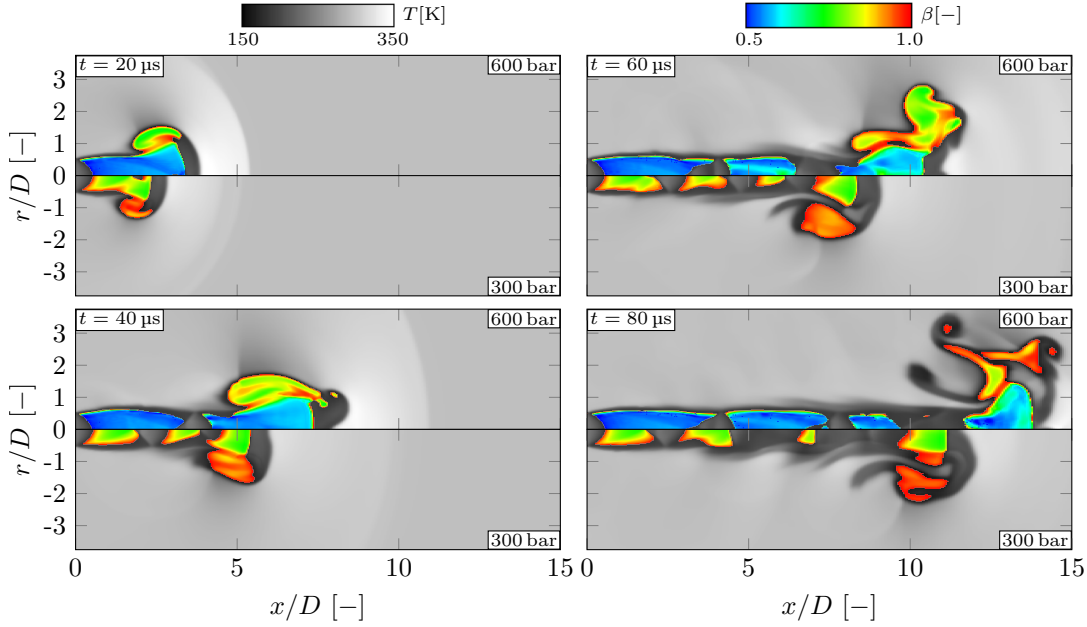}%

	\vspace{6pt}
	\caption{Comparison of the near-nozzle flow structure between the two CNG test-cases CNG-p600 (upper half) and CNG-p300 (lower half).}
	\label{fig:CNG_comparison}
	\vspace{12pt}
\end{figure}

In Fig.~\ref{fig:CNG_comparison} the near-nozzle flow structures of the direct injection of CNG for both total fuel pressures \SI{600}{\bar} and \SI{300}{\bar} are compared in terms of snapshots at different times after injection. The snapshots are a superposition of the temperature field with the corresponding vapor mass fraction. First of all, the overall flow-structure is very similar for both cases concerning fluid-dynamic and thermodynamic effects. For the case CNG-p300 a weaker expansion due to the lower pressure ratio of $\Pi=\SI{6}{}$ can be seen compared to the case CNG-p600. Nevertheless, the expansion process is still sufficient to lead to the instability of the single-phase and to enter the VLE. The weaker expansion manifests itself directly in a larger vapor mass fraction compared to the CNG-p600 case. In addition, the comparison shows that the traveling spherical vortex is also visible and shows phase separation effects, but in contrast to the high pressure jet the break-up is weakened in its intensity and spatial expansion. Due to the lower total fuel pressure and therefore a lower density and speed of sound at the nozzle exit, the penetration of the CNG-p300 jet is slower. Furthermore, the length of the stationary shock barrels is smaller by a factor of almost two, which correlates well with the predictions of Velikorody and Kudriakov~\cite{velikorodny2012numerical}.

\subsection{Effects of fuel change}

\begin{figure}[h]
	\centering
	\tikzsetnextfilename{CNGCHG300_comparison}%
	\input{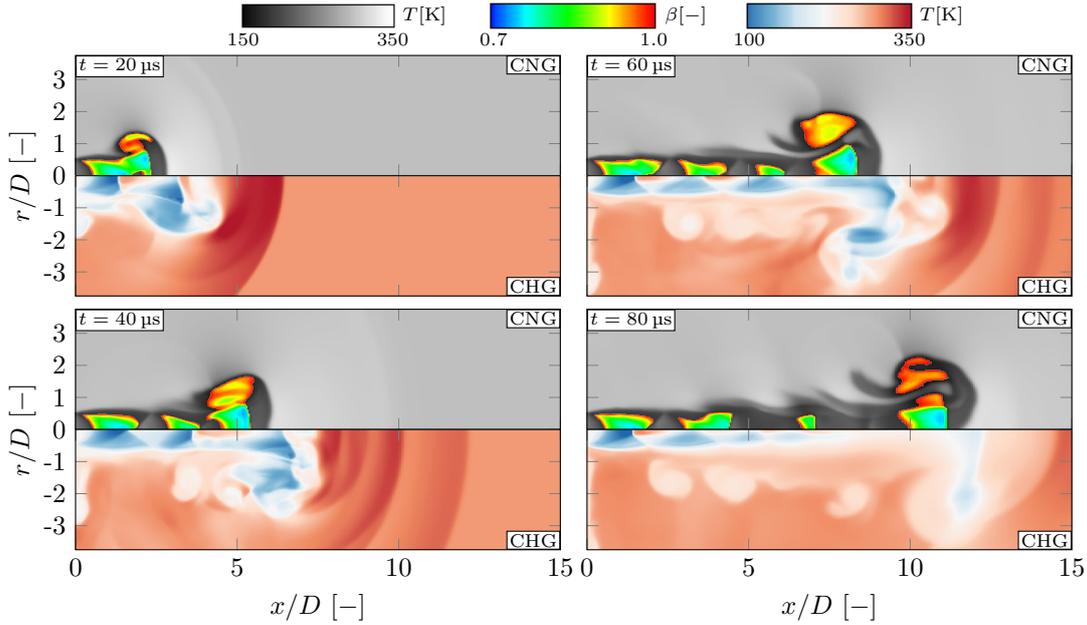}%

	\vspace{6pt}
	\caption{Comparison of the near-nozzle flow structure in terms of snapshots at four different times for CNG (upper half) and CHG (lower half) for a fuel pressure of \SI{300}{\bar}.}
	\label{fig:CNGCHG300_comparison}
\end{figure}

In the following we will compare the two different fuels, namely CNG and CHG, to each other. In this part we will focus on the 300 bar cases because the comparison of the 600 bar cases is almost identical. Figure~\ref{fig:CNGCHG300_comparison} shows the effects of the fuel change from CNG to CHG for a total pressure of \SI{300}{\bar} in terms of the temperature field superimposed by the vapor mass fraction, whereby the temperature ranges between 150 K and 350 K for the CNG-p300 case and between 100 K and 350 K for the CHG-p300 case. Both jets show the typical flow structures accompanying underexpanded jets but for the hydrogen case the shock and vortex structures are slightly more pronounced and show a larger spatial extent whereas the length of the shock barrels is a little bit shorter. Although the expansion of the CHG is much stronger (manifesting in a lower static temperature), the mixture remains in a single-phase state through out the complete jet. For the jet center this gets more clear when reconsidering the VLEs shown in Fig.~\ref{fig:VLEHydrogenNitrogen}. As nearly no dilution occurs in the jet core and the CHG mixture is almost pure hydrogen, see Tab~\ref{tab:CNGmixture}, an expansion to 100 K is by far not sufficient to enter the VLE of the binary mixture. In the region of the traveling spherical vortex the dilution of the fuel with the surrounding nitrogen would be sufficient but here in turn the expansion is too weak to enter the VLE.\\

\begin{figure}[h]
	\centering
	\tikzsetnextfilename{Ztip}%
	\input{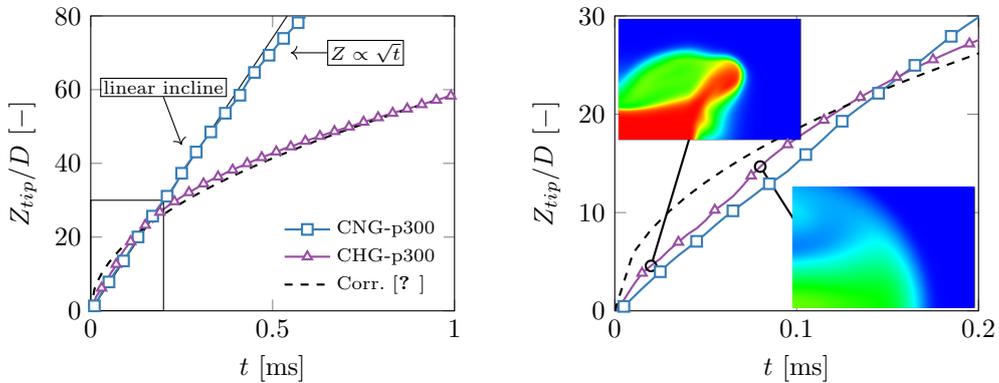}%

	\vspace{6pt}
	\caption{Left: Normalized jet penetration depth. Right: Detailed view of the injection profile up to \SI{0.2}{\milli\second}. Additional snapshots are plotted showing the detachment and reattachment of the jet for \SI{0.02}{\milli\second} and \SI{0.08}{\milli\second}, respectively. The contour plots are colored using the mass fraction $Y_{\textrm{CH}_4}$ ranging from \SI{0}{} (blue) to \SI{1}{} (red).}
	\label{fig:Ztip}
\end{figure}

In Fig.~\ref{fig:Ztip} the normalized jet penetration depth ${Z}_{tip}/D$ tracked with a mass fraction of $Y_{fuel}=\SI{0.01}{}$ at the centerline is shown for the two low-pressure test-cases CNG-p300 and CHG-p300. Up to \SI{0.2}{ms} the kinetic momentum flux $J^\ast = {a_s^\ast}^2 A$ is dominant \cite{abraham1996entrapment} and the CHG-p300-jet penetrates faster than the CNG-p300-jet with the same total pressure, see also Fig.~\ref{fig:CNGCHG300_comparison}. In the beginning of the injection the penetration curve of the CHG-jet shows a linear incline up to \SI{0.08}{\milli\second}. Comparing snapshots of the start and end time of the linear profile the reason for this behavior becomes obvious, being a detachment of the jet from the centerline, see Fig.~\ref{fig:CNGCHG300_comparison} on the right side. A blocking cone is formed in front of the jet consisting of a higher pressure and therefore higher density region which redirects the main flow away from the centerline to the sides, an effect which was also noticed for high pressure gas injections by other researchers, see, e.g., Hamzehloo \textit{et~al.}~\cite{hamzehloo2016gas}. After the cone vanishes the jet reattaches to the centerline and the injection profile is in a very good agreement with the correlation of Hill and Ouellette~\cite{hill1999transient} with the fitting parameter of $\Gamma=\SI{2.968}{}$ obtained in Banholzer \textit{et~al.}~\cite{banholzer2017}, see dashed lines in Fig.~\ref{fig:Ztip}. A first conclusion can be drawn that although the injection is performed at elevated pressures and therefore real-gas effect need to be modeled, the penetration depth for gas-like, single-phase jets can still be estimated by a rather simple correlation.

The results for the injection profile for the CNG-p300-case are completely different. Not only in the beginning, but almost throughout the whole computational domain the linear incline remains. Shortly before \SI{0.5}{\milli\second} it can be seen that the linear incline changes to the typical curvature known for transient gaseous jets and the jet penetration is proportional to the square root of time. One of the major reasons for this extraordinary behavior is the traveling spherical vortex. Fig.~\ref{fig:CNG_comparison} and \ref{fig:CNGCHG300_comparison} show large two-phase regions in the first vortex, which remain present even for longer simulation times. Later, increasing temperatures and pressures lead to the disappearance of those, momentum gets lost and the jet slows down. The time of vanishing of the two-phase region and the time where the slope changes its characteristic is identical. Further studies are necessary for a full understanding of the injection characteristics in high-pressure two-phase fuel injections.



\section{Conclusion and Outlook}
\label{sec:conclusion}

In order to simulate and discuss real-gas effects and phase separation processes occurring in underexpanded jets at engine-relevant conditions two major numerical challenges were addressed in this work. On the one hand, it is indispensable to derive numerical schemes which are capable of solving highly compressible flows including accurate predictions of shock- and flow-discontinuities. On the other hand, a consistent, appropriate and efficient modeling of the thermodynamic state considering real-gas effects and phase instability is needed. Hence, a numerical framework implemented in the open-source tool OpenFOAM is presented in this work combining a hybrid, pressure-based solver with a vapor-liquid equilibrium (VLE) model based on the cubic equation of state (EoS). This framework is used to investigate underexpanded jets at engine-relevant conditions. To the authors knowledge such an investigation is carried out for the first time in literature and therefore both numerical as well as experimental studies are lacking in this field. As a consequence, we strongly focused on a thorough validation of both parts of the numerical framework (solver and thermodynamics), whereby the validation of the solver was recently carried out by Kraposhin \textit{et~al.}~\cite{kraposhinbanholzer2017} by means of one-dimensional flow problems. Therefore, in the present study the focus was mainly on the validation of the thermodynamics employing general thermodynamic relations and measurement data available in the literature. In this context, the range of applicability of two different EoSs was discussed and the most important parts of the VLE-model were described. Additionally, compressibility effects and the variation of the speed of sound in the VLE-region was examined thoroughly.  A large number of measurement data was used to discuss the prediction accuracy of the VLE-model. Besides, a recent (experimental and numerical) investigation~\cite{traxinger2017a} about mixture induced phase separation for high-pressure injection in the low Mach number regime was simulated to validate the numerical framework set up in OpenFOAM. In this sense, three simulations were carried out where n-hexane is injected into nitrogen at elevated pressure. Based on all these successful validations, engine-relevant simulation cases for two different fuels (CNG and CHG) were defined with total pressures up to \SI{600}{\bar} and pressure ratios up to \SI{12}{}. While CHG showed no phase separation effects for the considered high-pressure injections, two-phase regions for the CNG-jets evolved. Analyses revealed that the phase separation are caused by two different effects. Firstly, by the strong expansion due to the large pressure ratio and secondly, by the mixing of the fuel with the chamber gas. A comparison of the single-phase with the two-phase jets disclosed that the phase separation leads to a completely different penetration depth in contrast to single-phase injection and therefore commonly used analytical approaches fail to predict the penetration depth.

In future studies several different topics have to be addressed. Different fluids and injection conditions additional to the ones presented in this work are important to be investigated in order to reveal further conditions where phase separation may occur. In this context, the difference in penetration depth between single- and two-phase jets has to be further examined. Possible non-equilibrium effects in terms of phase separation have to be inquired as they can delay phase separation especially in high Mach number jets as it is for instance the case in last stages of low-pressure steam turbines. 

\section*{Acknowledgements}

The authors would like to thank Munich Aerospace (\textit{www.munich-aerospace.de}) and The Research Association for Combustion Engines eV (FVV, Frankfurt, \textit{www.fvv-net.de}) for the funding. Furthermore, we acknowledge the Dortmund Data Bank Software \& Separation Technology (DDBST) for their support in finding appropriate measurement data for the investigated binary and ternary mixtures.

\bibliographystyle{unsrt}
\bibliography{bibtex.bib}

\end{document}